\documentclass[aps,prb,reprint,twocolumn,floatfix,superscriptaddress,longbibliography]{revtex4-2}

\usepackage{standalone}
\usepackage{amsthm}
\usepackage{lipsum}
\usepackage{amsfonts, amsmath, amssymb}
\usepackage{graphicx}
\usepackage{dcolumn}
\usepackage{bm}
\usepackage{dsfont}
\usepackage[normalem]{ulem}
\usepackage{color}
\usepackage[utf8]{inputenc}
\usepackage[colorlinks,citecolor=blue,linkcolor=blue,urlcolor=blue]{hyperref}
\usepackage{tikz}
\usepackage{braket}
\usepackage{physics}
\usepackage{color}
\usepackage{soul}
\usepackage{mathtools}
\usepackage{float}
\usepackage{esvect}
\usepackage{subfigure}
\usepackage{stmaryrd} 
\usepackage{enumerate}

\include{physics}
\begin{document}

\title{Quantum electron transport controlled by cavity vacuum fields}

\author{Geva Arwas}
\affiliation{Universit\'{e} Paris Cit\'e, CNRS, Mat\'{e}riaux et Ph\'{e}nom\`{e}nes Quantiques, 75013 Paris, France}

\author{ Cristiano Ciuti}
\affiliation{Universit\'{e} Paris Cit\'e, CNRS, Mat\'{e}riaux et Ph\'{e}nom\`{e}nes Quantiques, 75013 Paris, France}

\begin{abstract}
\noindent
    We explore theoretically how the coupling to cavity vacuum fields affects the electron transport in quantum conductors  due to the counter-rotating-wave terms of light-matter interaction. We determine the quantum conductance in terms of the transmission coefficients predicted by an effective electron Hamiltonian. The coupling between bare electronic states is mediated by virtual processes involving intermediate states with one electron (or one hole) on top of the Fermi sea and one virtual cavity photon.
    We study the behavior of the quantum conductance in the presence of artificial or disordered single-particle potentials, as well as a spatially varying cavity mode. As illustrative examples, we apply our theory to 1D conductors and to disordered 2D quantum Hall systems.   We show how the cavity vacuum fields can lead to both large enhancement or suppression of electron conductance in the ballistic regime, as well as modification of the conductance quantization and fluctuations.
\end{abstract}
\date{\today}
\maketitle

\section{Introduction}
The manipulation of matter by the coupling to cavity fields is an emerging topic in physics and chemistry \cite{genet2021inducing,garcia2021manipulating,bloch2022strongly,Schlawin2022}. An aspect of particular interest is the control of electronic properties by a passive electromagnetic resonator, i.e., in the absence of illumination \cite{Nataf2019,Schlawin2019,Eckstein2020,Guerci2020,Andolina2020,Amelio2021}. {\color{black}The control of electron transport by vacuum fields is one of the open frontiers \cite{orgiu2015conductivity,nagarajan2020conductivity,ParaviciniBagliani2018,Appugliese2022,Valmorra2021}.  Experiments on disordered organic materials have reported a conductance enhancement without illumination, mediated by the presence of cavity photon modes that are spatially extended over the whole electronic system  \cite{orgiu2015conductivity,nagarajan2020conductivity}}. Early theoretical studies have explored this problem via tight-binding models dealing with chains of two-level systems describing the hopping of electrons between different molecular orbitals in the organic material, accounting both for disorder and light-matter interaction in the rotating-wave approximation \cite{hagenmuller2017cavity,Hagen_PRB_2018,Chavez2021}. In these models, however, electrons are supposed to be injected in the excited band that is separated from the ground band by an energy close to that of the cavity photon: in this scenario, real photons can be emitted by spontaneous emission and subsequently resonantly re-absorbed via the standard rotating-wave terms of light-matter interaction in a dynamics that is related to cavity electro-luminescent devices \cite{DeLib2008,DeLib2009}.

Yet, without illumination, for vanishing applied voltage and low temperature, electrons cannot be injected into highly excited states and real photons cannot be emitted. Earlier works in the framework of linear-response Kubo formalism did not consider disorder, but only scattering at the phenomenological semi-classical level \cite{Bartolo_2018,ParaviciniBagliani2018,NaudetBaulieu2019}, where the response is mediated by collective polariton excitations \cite{Paulillo:17,Jeannin2019} in the ultrastrong light-matter coupling regime \cite{FriskKockum2019,RevModPhys.91.025005}. In a spatially in-homogeneous system, a possible mechanism for electron hopping to be affected by vacuum fields is the exchange of virtual cavity photons via the counter-rotating terms of light-matter interaction, which has been recently introduced for disordered quantum Hall systems \cite{Ciuti2021}. Such processes can in principle mediate long-range electron hopping affecting the edge channel transport and breaking down the topological protection of the integer quantum Hall effect, as discovered experimentally \cite{Appugliese2022}.

In the general context of quantum conductors, a microscopic theory of electron transport mediated by cavity vacuum fields that takes into account the effect of counter-rotating-wave terms of light-matter interaction, mode field spatial profile and arbitrary single-particle electronic potential is still missing. In this article, we address this general problem. To achieve that, we determine a cavity-mediated effective Hamiltonian for a Fermi system that can be conveniently exploited to predict transmission coefficients from one probe lead to another one. We show how, depending on the single-particle electron potentials and spatial mode profile, the cavity vacuum fields can both dramatically enhance or suppress the quantum electron conductance, as well as strongly affecting its quantization and fluctuation properties.

\section{Quantum light-matter Hamiltonian}
\begin{figure}[t!]
	\includegraphics[width=1\hsize]{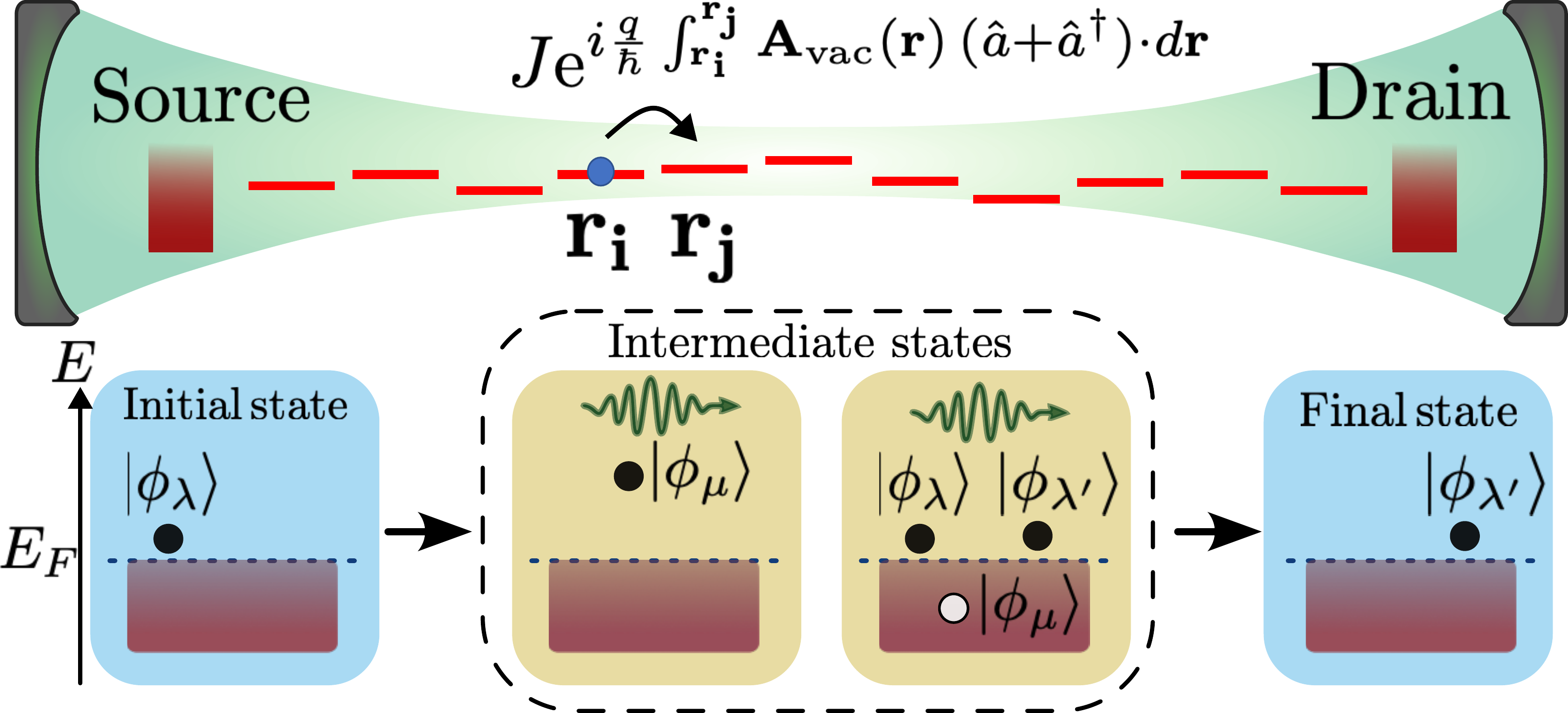}
	\caption{Top panel: sketch of the considered electronic system, described by a single-band tight-binding lattice model with hopping $J$ between nearest neighbors. The goal is to study how the linear quantum electron transport  is affected by the cavity vacuum fields (no illumination). Bottom panel: sketch of the process creating an effective coupling between single-particle states $\vert \phi_{\lambda} \rangle$ and $\vert \phi_{\lambda'}\rangle$ via two types of intermediate states consisting of one virtual cavity photon and one electron (or hole) in another state $\vert \phi_{\mu}\rangle$.\label{sketch}}
\end{figure}

Let us consider a single-band electronic conductor in a single-mode cavity resonator, as sketched in Fig. \ref{sketch}. Within a tight-binding description, the Hamiltonian can be written as:
%
\begin{equation} 
\hat{\mathcal{H}} =    \sum_{\mathbf{i}} E_{\mathbf{i}}  \hat{d}_{\mathbf{i}}^{\dagger }\hat{d}_{\mathbf{i}} 
   -\sum_{\langle \mathbf{i},\mathbf{j}\rangle  } 
   J \mathrm{e}^{i  \frac{q}{\hbar}  \int_{\mathbf{r}_\mathbf{i}}^{\mathbf{r}_\mathbf{j}} \hat{\mathbf{A}}(\mathbf{r})  \cdot d\mathbf{r} }  \hat{d}_{\mathbf{i}}^{\dagger }\hat{d}_{\mathbf{j}} 
      +   \hbar \omega_{\mathrm cav}  \hat{a}^{\dagger} \hat{a}   \, ,
\end{equation}
where $ \hat{d}_{\mathbf{i}}^{\dagger } $ creates an electron with charge $q$ in the lattice site labeled by $\mathbf{i}$ and position $\mathbf{r}_{\mathbf{i}}$, while $ E_\mathbf{i} = \epsilon + U(\mathbf{r}_{\mathbf i})$ is the on-site energy that can include a disordered and/or an arbitrary artificial single-particle potential $U(\mathbf{r})$. The nearest-neighbor hopping coupling is  $J$, while  $ \hat{\mathbf{A} }(\mathbf{r})
= \mathbf{A}_{\mathrm{vac}} (\mathbf{r}) \, (\hat{a}+ \hat{a}^{\dagger} )$ is the electromagnetic vector potential with vacuum amplitude $A_{\mathrm{vac}}(\mathbf{r})$.
The considered cavity mode has frequency $\omega_{\mathrm{cav}}$ and a bosonic creation (destruction) operator $\hat{a}^{\dagger}$ ($\hat{a}$). 
The quantum light-matter Hamiltonian can be rewritten as $\hat{\mathcal{H}} = \hat{\mathcal{H}}_0 + \hbar \omega_{\mathrm{cav}}   \hat{a}^{\dagger} \hat{a} + \hat{V}$, with 
$\hat{\mathcal{H}}_0$ the bare electronic Hamiltonian
\begin{equation} 
\hat{\mathcal{H}}_0  =    \sum_{\mathbf{i}} E_{\mathbf{i}}  \hat{d}_{\mathbf{i}}^{\dagger }\hat{d}_{\mathbf{i}} 
   -J\sum_{\langle \mathbf{i},\mathbf{j}\rangle  } \hat{d}_{\mathbf{i}}^{\dagger }\hat{d}_{\mathbf{j}}   
       = \sum_{\lambda} \epsilon_{\lambda} \hat{c}_{\lambda}^{\dagger }\hat{c}_{\lambda} \, , 
\end{equation}
where $\hat{c}_{\lambda}^{\dagger }$ is the fermionic creation operator for the single-particle eigenstate $\vert \phi_{\lambda} \rangle$ with energy $\epsilon_{\lambda}$. The quantum light-matter interaction is:
\begin{equation}
      \hat{V} = -J\sum_{\langle \mathbf{i},\mathbf{j}\rangle  } \sum_{n=1}^{\infty} \frac{1}{n!}
      (\mathrm{i} g_{\mathbf{ij}})^n (\hat{a}+ \hat{a}^{\dagger} )^n 
      \hat{d}_{\mathbf{i}}^{\dagger }\hat{d}_{\mathbf{j}} \, ,
\end{equation}
where $g_{\mathbf{ij}}=  \frac{q}{\hbar} \int_{{\mathbf r}_\mathbf{i}}^{\mathbf{r}_\mathbf{j}} \mathbf{A}_{\mathrm{vac}}({\mathbf r}) \,    \cdot d {\mathbf r}  $. The interaction $\hat{V}$ can be conveniently re-expressed in the basis of the single-particle eigenstates as follows:
\begin{eqnarray} 
\hat{V} &=& -J\sum_{\lambda,\lambda'  } \sum_{n=1}^{\infty} \frac{\mathrm{i}^n}{n!}
 \tilde{g}_{\lambda,\lambda'}^{(n)} (\hat{a}+ \hat{a}^{\dagger} )^n 
      \hat{c}_{\lambda}^{\dagger }\hat{c}_{\lambda'} \,  , \label{e4}
\end{eqnarray}
with 
\begin{equation} 
    \tilde{g}_{\lambda \lambda^{'}  }^{(n)} \ = \   \sum_{\langle \mathbf{i,j} \rangle  }
   g_{\mathbf{ij}}^n \langle \phi_{ \lambda} | \mathbf{i} \rangle    \langle \mathbf{j} | \phi_{ \lambda^{'} } \rangle \, .
\end{equation}
{\color{black} The light-matter interaction can be treated in a perturbative manner when the characteristic single-electron vacuum Rabi energy $|J \tilde{g}_{\lambda \lambda^{'}  }^{(1)} |$ is much smaller than the photon energy $\hbar \omega_{\mathrm{cav}} $. When $ |\tilde{g}_{\lambda \lambda^{'}  }^{(2)}| \ll  |\tilde{g}_{\lambda \lambda^{'}  }^{(1)} | $, then only the term $n = 1$ can be retained at the lowest-order in perturbation theory, which involve only process with one virtual cavity photon. For simplicity, we will denote $\tilde{g}_{\lambda \lambda^{'}  }^{(n=1)}$ with $\tilde{g}_{\lambda \lambda^{'}  }$.}


\section{Effective electronic Hamiltonian and conductance}
Our goal is to study the quantum electron transport with vanishing voltage bias at low temperatures. In this regime, no real photons can be emitted and the cavity can mediate interactions only via the exchange of virtual cavity photons.  Let us assume that the many-electron ground state is given by the Fermi sea $ |G \rangle =  \Pi_{\epsilon_\nu \leq E_\mathrm{F}} \hat{c}_\nu^\dagger | 0 \rangle$, where $E_\mathrm{F}$ is the Fermi energy. If we consider an electron injected above the Fermi level in a state $\hat{c}^\dagger_\lambda  | G \rangle$ with $\epsilon_\lambda \gtrsim E_\mathrm{F}$ and consider processes involving one virtual cavity photon, then the quantum light-matter interaction produces an effective coupling of $\hat{c}^\dagger_\lambda  | G \rangle$ to another state  $\hat{c}^\dagger_{\lambda'}  | G \rangle$ with $\epsilon_{\lambda'} \gtrsim E_\mathrm{F}$, as illustrated in Fig. \ref{sketch}. The first type of process has an intermediate state with a cavity photon and an electron above the Fermi energy: 
\begin{eqnarray} 
   \hat{c}^\dagger_\lambda  | G \rangle \  \rightarrow \ \hat{c}^\dagger_\mu \hat{a}^\dagger | G \rangle \  \rightarrow \ \hat{c}^\dagger_{\lambda^{'}}  | G \rangle \, ,
\end{eqnarray}
where $\epsilon_\mu \geq E_\mathrm{F}$. The product of the matrix elements of $\hat{V}$ for the two steps is $- J^2 \tilde{g}_{\lambda, \mu} \tilde{g}_{\mu, \lambda'}$. The energy difference (penalty) between the initial and intermediate state is $\epsilon_\lambda - \epsilon_\mu - \hbar \omega_{cav} < 0$.   
The second type involves an electron-hole excitation:
\begin{eqnarray} 
  \hat{c}^\dagger_\lambda  | G \rangle \  \rightarrow \    \hat{c}^\dagger_\lambda  \hat{c}^\dagger_{\lambda^{'}}     \hat{c}_\mu \hat{a}^\dagger  | G \rangle \  \rightarrow \   \hat{c}^\dagger_{\lambda^{'}} | G \rangle \, ,
\end{eqnarray}
where $\epsilon_\mu \leq E_\mathrm{F}$. The product of the matrix elements of $\hat{V}$ for the two steps is $+ J^2 \tilde{g}_{\mu,\lambda'} \tilde{g}_{\lambda, \mu}$ (the sign is different with respect to the first process due to fermionic anti-commutation). The energy difference between the initial and intermediate state is $\epsilon_{\mu} - \epsilon_{\lambda'} - \hbar \omega_{cav} <0$.
The effective coupling $\tilde{\Gamma}_{\lambda,\lambda'}$ between $\hat{c}^\dagger_\lambda  | G \rangle$ and $\hat{c}^\dagger_{\lambda'}  | G \rangle$, obtained by  elimination of the intermediate states  \cite{Malrieu_1985, Effective_PRB}, can be approximated as {\color{black} (see Appendix)}:
\begin{equation}
\label{Gamma}
\tilde{\Gamma}_{\lambda,\lambda'}\simeq  \sum_{\epsilon_{\mu} \geq E_F}   \frac{  J^2\, \tilde{g}_{\lambda,\mu} \tilde{g}_{\mu,\lambda'}}{ \epsilon_{\mu} - \epsilon_{\lambda}   + \hbar \omega_{cav}} 
- \sum_{\epsilon_{\mu} < E_F}  \frac{ J^2\, \tilde{g}_{\lambda,\mu} \tilde{g}_{\mu,\lambda'}}{ \epsilon_{\lambda'} - \epsilon_{\mu}   + \hbar \omega_{cav}}.
\end{equation}
To determine the quantum electron transport, we consider the effective single-particle Hamiltonian for the electronic states:
\begin{equation}
\hat{\mathcal H}_{\mathrm{eff}} = 
\label{H_eff}
\sum_{\lambda} \epsilon_{\lambda} \vert \phi_{\lambda} \rangle \langle \phi_{\lambda} \vert  +
\sum_{\lambda, \lambda' } 
\frac{\tilde{\Gamma}_{\lambda,\lambda'} +  
\tilde{\Gamma}^*_{\lambda',\lambda}}{2}\vert \phi_{\lambda} \rangle \langle \phi_{\lambda'} \vert \, .
\end{equation}
Given that we are interested in the quantum transport of electrons injected at the Fermi level, the relevant states are such that $\epsilon_{\lambda} \approx E_F$ and  $\epsilon_{\lambda'} \approx E_F$. Hence, we can approximate the sign function as $\mathrm{sgn}(\epsilon_{\mu}- E_F) \simeq \mathrm{sgn}(\epsilon_{\mu}- \mathrm{min}(\epsilon_{\lambda},\epsilon_{\lambda'}))$. Considering the different signs in the two sums in Eq. (\ref{Gamma}), this allows us to get the following compact approximation (see Appendix):
\begin{equation} \label{gammaTot}
\tilde{\Gamma}_{\lambda,\lambda'} \simeq  J^2 \sum_{\mu}  \frac{\mathrm{sgn}(\epsilon_{\mu}- \mathrm{min}(\epsilon_{\lambda},\epsilon_{\lambda'}))   \, \tilde{g}_{\lambda,\mu} \tilde{g}_{\mu,\lambda'} }{ |\epsilon_\mu - \frac{\epsilon_{\lambda}+\epsilon_{\lambda'}}{2} |  + \hbar \omega_{cav}} \, ,
\end{equation}
where $\mu$ here labels single-particle states with energy both below and above $E_F$. 
{\color{black} Using $\hat{\mathcal H}_{\mathrm{eff}}$ and  Green's function approach (see Appendix), we can determine the transmission coefficients $T_{j,j'}$ at the Fermi energy between source ${\mathcal S}$ and drain ${\mathcal D}$ contacts. Here, $j$ and $j'$ denote the channels (propagating modes) of the contacts, which are modeled by ideal leads. The corresponding quantum conductance \cite{Datta2005,girvin_yang_2019} reads:}
\begin{equation}
G(E_{\mathrm{F}}) = \frac{e^2}{h} \sum_{j \in {\mathcal S}\,, j' \in {\mathcal D}}  T_{j,j'}(E_{\mathrm{F}})  \, .    
\end{equation}
\begin{figure}[t!]
	\includegraphics[width=1\hsize]{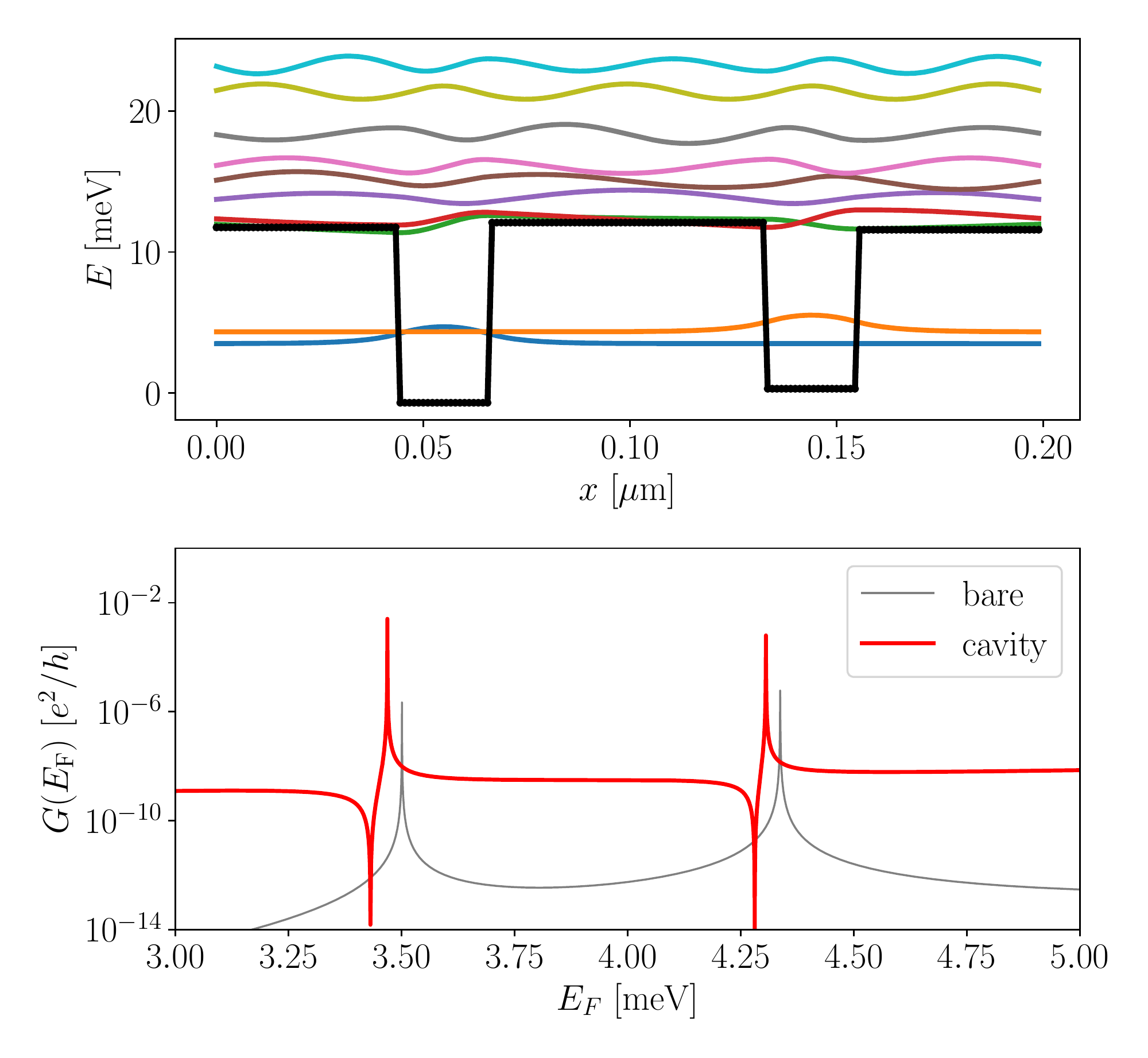}
	\caption{Example of conductance enhancement by cavity vacuum fields in a 1D system. The cavity mode is spatially homogeneous, linearly polarized along the $x$ direction, with frequency $\omega_{\mathrm{cav}}= 2 \pi \times 1 \mathrm{THz}$ and vacuum field $A_{\mathrm{vac}}$ that corresponds to  $\epsilon_r \eta = 10^{-9} $ (see text). Other parameters: $m = 0.067 m_{0}$ and $N=180$. Top panel: single-particle potential $U(x)$ (thick black line). The shapes of the wavefunctions for the energy eigenstates are also shown, vertically offset by their corresponding energies.  Bottom panel: conductance versus $E_F$ for the bare case (gray line) and for the cavity case (thick red line). 
		\label{double_enhancement}}
\end{figure}
\begin{figure}[t!]
\includegraphics[width=1\hsize]{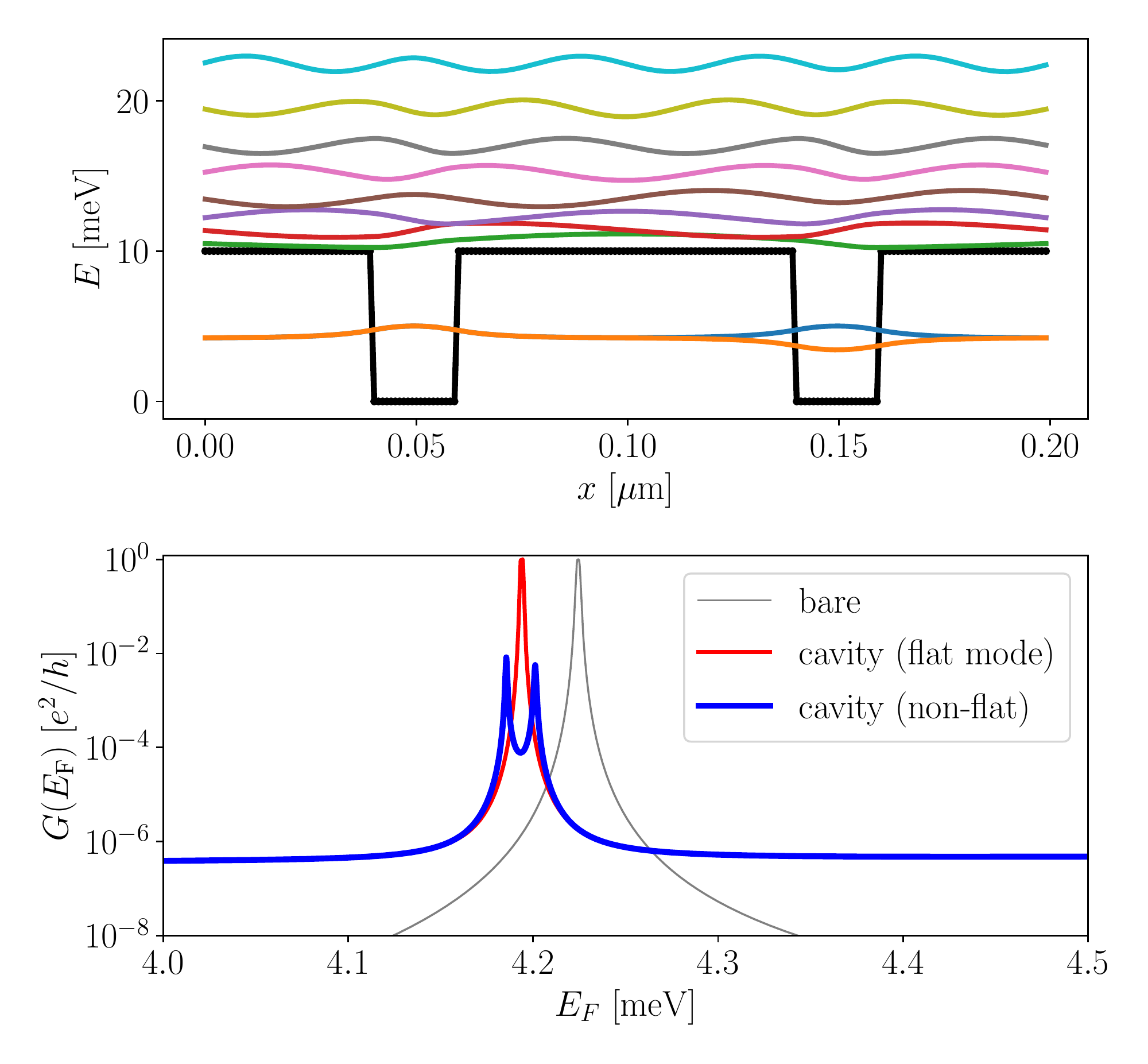}
\caption{Example of conductance suppression in a 1D system. Same plots and parameters as in Fig. \ref{double_enhancement} (except $N=200$), but with a symmetric double well. In the bottom panel, the red line corresponds to a flat cavity mode (as in Fig. \ref{double_enhancement}). The thicker blue line corresponds to a vacuum field $A_{\mathrm{vac}}(x) $ varying linearly from  $0.75 A_0$ to $ 1.25 A_0$ where $A_0$ corresponds to the same field as in Fig. \ref{double_enhancement}. 	\label{double_suppression} }
\end{figure}
\section{Results for 1D systems}
As a first application, we will consider a one-dimensional (1D) conductor with an artificial single-particle potential. This kind of systems can be engineered for example by using gates to induce tailored potentials on a 2D electron gas \cite{QPC_1988,wharam1988one,kane1998quantized,irber2017quantum}. In Fig. \ref{double_enhancement}, we consider an asymmetric double well and a spatially homogeneous cavity mode. For a 'flat` mode and a 1D lattice with length $L$ and $N$ sites, we have 
$g_{\mathbf{i j}}= g_{\mathrm{flat}}= \frac{\omega_{\mathrm{cav}}}{2 \pi c} \sqrt{\frac{\alpha}{\epsilon_r \eta}} d$, where $d = L/N$, $\alpha$ is the fine structure constant, $\epsilon_r$ the relative dielectric constant and $\eta$ is the compression factor of the cavity mode volume. This is defined by  $V_{\mathrm{mode}} = \eta \lambda_0^3$ with $\lambda_0 = 2 \pi c / \omega_{\mathrm{cav}}$ the free space wavelength. For split-ring resonators such compression factor $\eta$ can be as small as $\sim 10^{-10}$ \cite{keller2017few}. The hopping can be expressed as $J = \hbar^2/(2 m d^2)$, where $m$ is the band effective mass. For a fixed length $L$ and a sufficiently large $N$, the tight-binding model converges to its continuum limit. 
We remark that in the continuum limit, only the 1st and 2nd order terms of Eq. \ref{e4} tend to a well-defined value, while all higher order terms vanish for any interaction strength. Specifically, the first and second order terms in the expansion correspond to the paramagnetic and diamagnetic terms of the continuum minimal coupling Hamiltonian.

As shown in Fig. \ref{double_enhancement}, each well has a bound state. Without the cavity, these states have a small hybridization due to the asymmetric potential and relatively thick separation barrier.  As a result, the conductance (gray line) shown in the bottom panel of Fig. \ref{double_enhancement} displays two resonances, but with very low peak values in units of the conductance quantum $e^2/h$ due to the poor transmission ($\sim 10^{-6}$). The vacuum fields can mediate an effective interaction between these two bound states and enhance the tunnel transport. Indeed, as shown by the thick red line, the conductance exhibits an enhancement by several orders of magnitude. Moreover,  the enhanced peaks exhibit a Fano-like lineshape. This is due to the cavity-mediated coupling of the bound states to the unbound states. Indeed, we have verified that by neglecting such coupling, a Lorentzian lineshape is recovered. 

While the example in Fig. \ref{double_enhancement} displays a strong conductance enhancement, a different configuration can lead to a dramatic suppression (Fig. \ref{double_suppression}). The parameters are the same as in Fig. \ref{double_enhancement}, but  with a symmetric potential. Without cavity, the bound states in the two wells hybridize leading to a conductance resonance peak close to the conductance quantum. Due to the finite barrier, the transmission resonance has an intrinsic finite broadening, masking the small energy splitting between the two symmetric and anti-symmetric electron wavefunctions. Now, if we consider the same cavity flat mode as in Fig. \ref{double_enhancement}, then the effect of the vacuum fields is just a shift of the conductance resonance (thick red line). However, if the cavity mode has a moderate linear variation along the $x$ direction (from $0.75$ to $1.25$ times the average value), then the cavity-mediated coupling breaks the symmetry and the hybridization of the two bound states is significantly reduced. As a result, the conductance is dramatically suppressed by several orders of magnitude (two peaks can now be resolved, thicker blue line). In contrast, for the asymmetric electronic potential in Fig. \ref{double_enhancement}, the same spatial gradient of the cavity mode produces only a very minor modification of the conductance {\color{black} (see appendix)}.
\begin{figure}[t!]
\includegraphics[width=1\hsize]{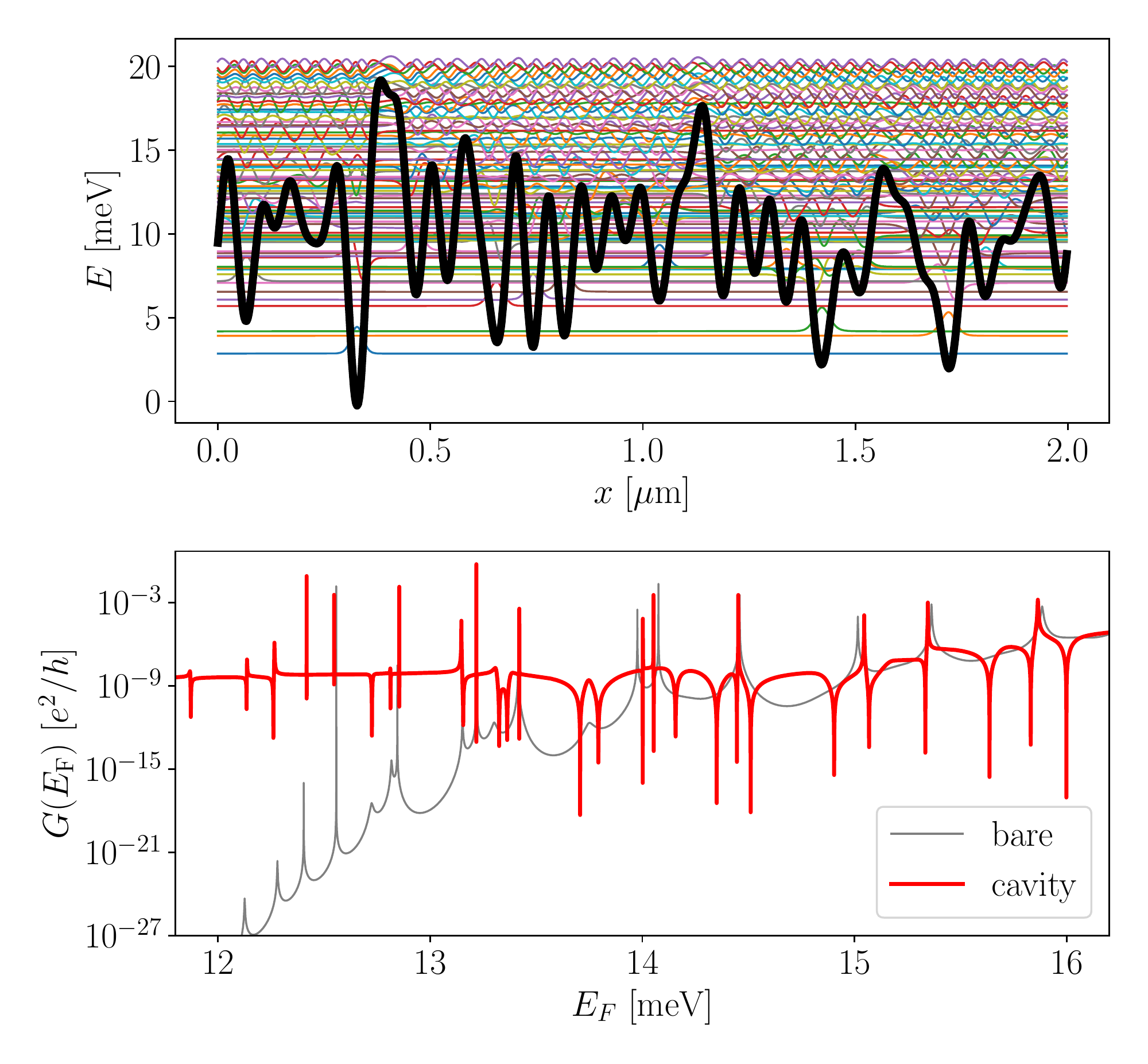}
\caption{Same plots and parameters as in Fig. \ref{double_enhancement} (except $N=800$), but with a smooth random potential $U(x)$ with a finite correlation length (thick black line) and with a spatially flat cavity mode.\label{disorder}}
\end{figure}
Now, let us explore the role of the cavity vacuum fields on a more complex disordered potential. As shown in Fig. \ref{disorder}, we have considered a smooth potential with a finite correlation length and displaying a significant number of peaks and valleys. At low energies, there are bound states trapped in the deepest wells. By increasing the energy, the probability of tunneling between different wells increases. At high energy a quasi-continuum spectrum of levels eventually emerges. The bottom panel of Fig. \ref{disorder} presents the conductance as a function of $E_F$ for the intermediate spectral region. Without the cavity, the conductance displays several resonance peaks. Increasing the energy, the envelope of the bare conductance significantly increases. With the considered cavity coupling, the conductance associated to the lowest energy peaks is enhanced by several orders of magnitude and acquires the characteristic Fano lineshape previously discussed. Note that some conductance peaks are suppressed.
\begin{figure}[t!]
\includegraphics[width=1\hsize]{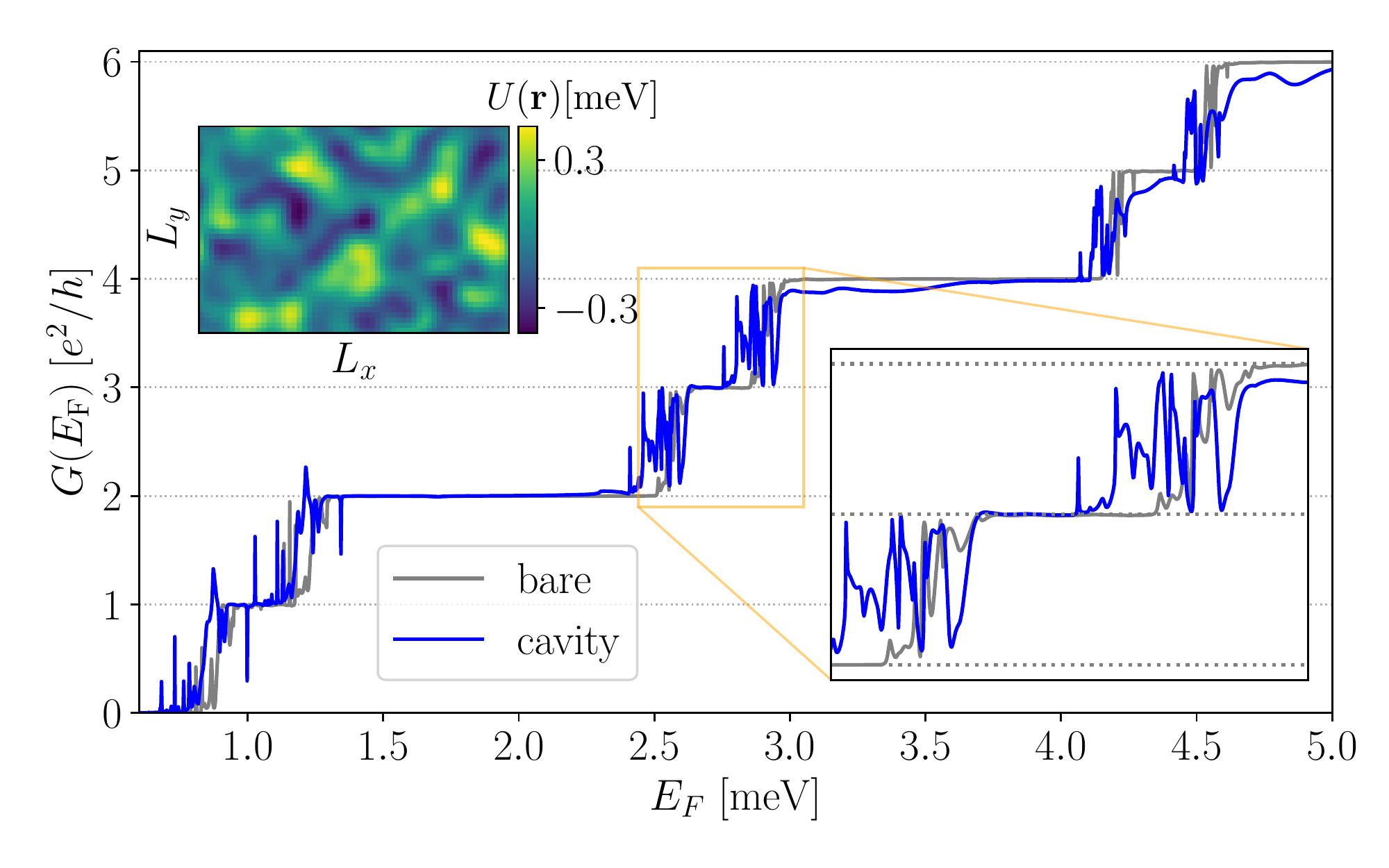}
\caption{Two-terminal conductance for a 2D system ($N = 60 \times 40$) with a perpendicular magnetic field $B$ for the bare (gray line) and cavity (blue line) case. The current flows along the $x$ direction. Parameters: $B = 1 \mathrm{Tesla}$, $m = 0.067 m_0$, $L_x = 0.5 \mathrm{\mu m}$, $L_y/L_x = 2/3$, $E_{\mathrm{Zeeman}} = 0.2 \hbar \omega_{\mathrm{cyc}}$ with $\omega_{\mathrm{cyc}} = e B/m$ the cyclotron frequency, and $\omega_{\mathrm{cav}}= 2 \pi \times 1 \mathrm{THz}$. The vacuum field is polarized along the $y$ direction and its amplitude $A_{\mathrm{vac}}(y) $ varies linearly from  $0.5 \bar{A}$ to $ 1.5 \bar{A}$, where the average $\bar{A}$ corresponds to $\epsilon_r \eta = 4 \times 10^{-10}$. The considered smooth 2D random potential $U(\mathbf{r})$ with a finite correlation length is shown in the inset.
\label{QH}}
\end{figure}

\section{Results for 2D systems}
Another interesting example is the quantum transport in a 2D conductor subject to a perpendicular magnetic field $B$. As shown in Fig. \ref{QH}, we have calculated the two-terminal conductance of a rectangular sample. We have included two spin channels (light-matter interaction conserves spin) with a finite Zeeman energy splitting. Moreover, we have considered a smooth 2D disordered potential and a cavity mode with a spatial gradient (see caption). In the bare case  (gray curve), by increasing the Fermi energy we can see clear quantized plateaus of the conductance at multiple integers of $e^2/h$. In the two-terminal configuration, this is due to the occupation of 1D edge states that cannot back-scatter \cite{Buttiker1988,Baranger1989} even in the presence of disorder due to the well-known topological protection of quantum Hall systems \cite{girvin_yang_2019}. With cavity coupling, disorder and spatial gradient of the cavity mode, cavity-mediated hopping becomes possible \cite{Ciuti2021}. Our microscopic transport calculations \footnote{With the cavity-mediated effective Hamiltonian,  the calculation is heavier because the  matrix $\tilde{\Gamma}_{\lambda,\lambda'}$  is not sparse and describes an all-to-all coupling between the bare eigenstates. Correspondingly, also the calculation of the transmission coefficients is more demanding.} show that increasing the number of active edge channels (by increasing $E_F$) the conductance quantization worsens with the corresponding plateaus even completely destroyed, as observed in \cite{Appugliese2022}. Note also that in the cavity case the disorder-induced fluctuations of the conductance are significantly enhanced. 

In conclusion, we have shown how cavity vacuum fields can affect quantum electron transport. With the framework introduced here, it is possible to explore the interplay of cavity vacuum fields with different types of disorder,  weak links, multi-probe geometries in a wide variety of configurations. The present theory with a single-band conductor and arbitrary single-particle potential can be generalized to multi-band systems. Given that in mesoscopic systems the use of metallic gates with different geometries is common and can produce strong vacuum fields, their role in general cannot be overlooked. Possible intriguing future investigations include the control of transport in 2D materials with non-trivial single-particle states such as van der Waals materials \cite{dean2013hofstadter,Rubio2022}, vertical semiconductor heterostructures  \cite{vigneron2019quantum,Limbacher2020} and hybrid semiconductor-superconductor systems \cite{hatefipour2021induced}. 

\acknowledgements
{We thank Danh-Phuong Nguyen for helpful discussions. We acknowledge support from the Israeli Council for Higher Education - VATAT, from FET FLAGSHIP Project PhoQuS (grant agreement ID no.820392) and from the French agency ANR through the project NOMOS (ANR-18-CE24-0026), TRIANGLE (ANR-20-CE47-0011) and CaVdW (ANR-21-CE30-0056-01).}

\appendix
\section{Derivation of the cavity mediated coupling}

In this section we present a detailed derivation of the cavity mediated coupling.
For the calculation of $\tilde{\Gamma}_{\lambda,\lambda'} $, we use the perturbative expansion introduced in \cite{Malrieu_1985}.
The effective coupling between an initial state $ | \Psi_\mathrm{I} \rangle $ and a final state $ | \Psi_\mathrm{F} \rangle $, due to the perturbation $  \hat{V} $, that is mediated by a set of intermediate states $| \Psi_\mathrm{int} \rangle $, can be written in the following form:
\begin{equation}
\label{SMeq1}
\Gamma_{\mathrm{IF}} =  \sum_{\mathrm{int}}   \frac{ \langle \Psi_\mathrm{I} |   \hat{V} | \Psi_\mathrm{int}   \rangle  \langle \Psi_\mathrm{int}  |  \hat{V}   | \Psi_\mathrm{F} \rangle    }{E_{\mathrm{I}} - E_{\mathrm{int}}  }, 
\end{equation}
where $E_{\mathrm{I}} - E_{\mathrm{int}} $ is the energy difference (penalty) between the initial and intermediate state.
In the present work, we are interested in the effective coupling between the initial state $\hat{c}^\dagger_{\lambda}  | G \rangle $ and the final state $\hat{c}^\dagger_{\lambda'}  | G \rangle $, where  $| G \rangle  $ is the many-electron ground state, given by the Fermi sea $ |G \rangle =  \Pi_{\epsilon_\nu \leq E_\mathrm{F}} \hat{c}_\nu^\dagger | 0 \rangle$, while $ \hat{c}^\dagger_{\lambda}  $ and $\hat{c}^\dagger_{\lambda'}$ are the electron creation operators for single-particle states with energy just above the Fermi level, namely $\epsilon_{\lambda} , \epsilon_{\lambda'} \gtrsim E_\mathrm{F}$. As illustrated in Fig. \ref{sketch}, we consider two types of intermediate states that involve a virtual cavity photon. The total effective coupling $\tilde{\Gamma}_{\lambda,\lambda'}$ is given by the sum of all processes.

The first type of process has an intermediate state with a cavity photon and an electron above the Fermi energy: 
\begin{widetext}
\begin{eqnarray} 
 \mathrm{Type \ I:} \ \ \ \ \  | \Psi_\mathrm{I} \rangle = \hat{c}^\dagger_\lambda  | G \rangle \  \longrightarrow \  | \Psi_\mathrm{int}   \rangle  = \hat{c}^\dagger_\mu \hat{a}^\dagger | G \rangle \  \longrightarrow \ | \Psi_\mathrm{F} \rangle = \hat{c}^\dagger_{\lambda^{'}}  | G \rangle \, ,
\end{eqnarray}
where $\epsilon_\mu \geq E_\mathrm{F}$. Using the perturbation $\hat{V}$ given by Eq. (\ref{e4}) to first order, we calculate the following matrix elements:
\begin{eqnarray} 
 \langle \Psi_\mathrm{I} |   \hat{V} | \Psi_\mathrm{int}   \rangle \ &=& \ 
  \langle G | \hat{c}_\lambda  \hat{V}  \hat{c}^\dagger_\mu \hat{a}^\dagger | G \rangle \ = \ 
  - \mathrm{i} J  \langle G | \hat{c}_\lambda  \tilde{g}_{\lambda,\mu} (\hat{a}+ \hat{a}^{\dagger} )  \hat{c}_{\lambda}^{\dagger }\hat{c}_{\mu}  \hat{c}^\dagger_\mu \hat{a}^\dagger | G \rangle \ = \ - \mathrm{i} J \tilde{g}_{\lambda,\mu}   
  \\
 \langle \Psi_\mathrm{int}  |  \hat{V}   | \Psi_\mathrm{F} \rangle  \ &=& \ 
  \langle G | \hat{a}  \hat{c}_\mu  \hat{V}   \hat{c}^\dagger_{\lambda^{'}}  | G \rangle \ = \ 
  - \mathrm{i} J   \langle G | \hat{a}  \hat{c}_\mu   \tilde{g}_{\mu,\lambda^{'}} (\hat{a}+ \hat{a}^{\dagger} )  \hat{c}_{\mu}^{\dagger } \hat{c}_{\lambda^{'}}    \hat{c}^\dagger_{\lambda^{'}}  | G \rangle \ = \ - \mathrm{i} J \tilde{g}_{\mu,\lambda^{'}} 
\end{eqnarray}
\end{widetext}
The energy difference (penalty) between the initial and intermediate state is $\epsilon_\lambda - \epsilon_\mu - \hbar \omega_{cav} < 0$. Using Eq. (\ref{SMeq1}), we can now write the cavity mediated hopping induced by the first process as:
\begin{equation}
\tilde{\Gamma}^{(\mathrm{Type \ I})}_{\lambda,\lambda'} \ = \  \sum_{\epsilon_{\mu} \geq E_F}   \frac{  J^2\, \tilde{g}_{\lambda,\mu} \tilde{g}_{\mu,\lambda'}}{ \epsilon_{\mu} - \epsilon_{\lambda}   + \hbar \omega_{cav}} .
\end{equation}

We now consider the second type of process, which involves electron-hole excitations. 
Unlike the first type of process, here the diagonal $\tilde{\Gamma}_{\lambda,\lambda}$ corrections involve different sets of intermediate states, which we address later on.
The main (off-diagonal) process is:
\begin{widetext}
\begin{eqnarray} 
 \mathrm{Type \ II:} \ \ \ \ \  | \Psi_\mathrm{I} \rangle = \hat{c}^\dagger_\lambda  | G \rangle \  \longrightarrow \     | \Psi_\mathrm{int}   \rangle  = \hat{c}^\dagger_\lambda  \hat{c}^\dagger_{\lambda^{'}}     \hat{c}_\mu \hat{a}^\dagger  | G \rangle \  \longrightarrow \  | \Psi_\mathrm{F} \rangle =  \hat{c}^\dagger_{\lambda^{'}} | G \rangle \, ,
\end{eqnarray}
where $\epsilon_\mu \leq E_\mathrm{F}$. Note that this process vanishes for $\lambda = \lambda'$ due to Pauli blocking.
Similarly to the first process, we calculate the following matrix elements:
\begin{eqnarray} 
 \langle \Psi_\mathrm{I} |   \hat{V} | \Psi_\mathrm{int}   \rangle \ &=& \ 
  \langle G | \hat{c}_\lambda  \hat{V}  \hat{c}^\dagger_\lambda  \hat{c}^\dagger_{\lambda^{'}}     \hat{c}_\mu \hat{a}^\dagger  | G \rangle \ = \ 
   - \mathrm{i} J  \langle G | \hat{c}_\lambda 
    \tilde{g}_{\mu,\lambda^{'}} (\hat{a}+ \hat{a}^{\dagger} )  \hat{c}_{\mu}^{\dagger } \hat{c}_{\lambda^{'}}
  \hat{c}^\dagger_\lambda  \hat{c}^\dagger_{\lambda^{'}}     \hat{c}_\mu \hat{a}^\dagger  | G \rangle \ = \
  - \mathrm{i} J \tilde{g}_{\mu,\lambda^{'}} \, ,
  \\
 \langle \Psi_\mathrm{int}  |  \hat{V}   | \Psi_\mathrm{F} \rangle  \ &=& \ 
  \langle G | \hat{a}  \hat{c}_{\mu}^{\dagger } \hat{c}_{\lambda^{'}}   \hat{c}_\lambda   \hat{V} \hat{c}^\dagger_{\lambda^{'}} | G \rangle \ = \ 
   - \mathrm{i} J   \langle G | \hat{a}  \hat{c}_{\mu}^{\dagger } \hat{c}_{\lambda^{'}}   \hat{c}_\lambda      \tilde{g}_{\lambda,\mu} (\hat{a}+ \hat{a}^{\dagger} )  \hat{c}_{\lambda}^{\dagger } \hat{c}_{\mu}         \hat{c}^\dagger_{\lambda^{'}} | G \rangle
  \ = \
   \mathrm{i} J  \tilde{g}_{\lambda,\mu} \, .
\end{eqnarray}
\end{widetext}
Note the sign difference between the second and first process, due to fermionic anti-commutation.
The energy difference between the initial and intermediate state is $\epsilon_{\mu} - \epsilon_{\lambda'} - \hbar \omega_{cav} <0$. We can now write the cavity-mediated hopping induced by the second process as:
\begin{equation}
\tilde{\Gamma}^{(\mathrm{Type \ II})}_{\lambda,\lambda'} \ = \ - (1 - \delta_{\lambda,\lambda'})  \sum_{\epsilon_{\mu} < E_F}  \frac{ J^2\, \tilde{g}_{\lambda,\mu} \tilde{g}_{\mu,\lambda'}}{ \epsilon_{\lambda'} - \epsilon_{\mu}   + \hbar \omega_{cav}}.
\end{equation}
The prefactor $ (1 - \delta_{\lambda,\lambda'}) $ is due to Pauli blocking and ensures that the diagonal elements vanish. Instead, the electron-hole contribution to the diagonal corrections is given by the following process:
\begin{widetext}
\begin{eqnarray} 
 \mathrm{Type \ II \ (diagonal):} \ \ \ \ \  | \Psi_\mathrm{I} \rangle = \hat{c}^\dagger_\lambda  | G \rangle \  \longrightarrow \     | \Psi_\mathrm{int}   \rangle  = \hat{c}^\dagger_\lambda  \hat{c}^\dagger_{\nu}     \hat{c}_\mu \hat{a}^\dagger  | G \rangle \  \longrightarrow \  | \Psi_\mathrm{F} \rangle =  \hat{c}^\dagger_{\lambda} | G \rangle \, ,
\end{eqnarray}
where $\epsilon_\mu \leq E_\mathrm{F}$, $\epsilon_\nu \geq E_\mathrm{F}$. The matrix elements for this diagonal process are:
\begin{eqnarray} 
 \langle \Psi_\mathrm{I} |   \hat{V} | \Psi_\mathrm{int}   \rangle \ &=& \ 
  \langle G | \hat{c}_\lambda  \hat{V}  \hat{c}^\dagger_\lambda  \hat{c}^\dagger_{\nu}     \hat{c}_\mu \hat{a}^\dagger  | G \rangle \ = \ 
   - \mathrm{i} J  \langle G | \hat{c}_\lambda 
    \tilde{g}_{\mu,\nu} (\hat{a}+ \hat{a}^{\dagger} )  \hat{c}_{\mu}^{\dagger } \hat{c}_{\nu}
  \hat{c}^\dagger_\lambda  \hat{c}^\dagger_{\nu}     \hat{c}_\mu \hat{a}^\dagger  | G \rangle \ = \
  - \mathrm{i} J \tilde{g}_{\mu,\nu}   \, ,
  \\
 \langle \Psi_\mathrm{int}  |  \hat{V}   | \Psi_\mathrm{F} \rangle  \ &=& \ 
  \langle G | \hat{a}  \hat{c}_{\mu}^{\dagger } \hat{c}_{\nu}   \hat{c}_\lambda   \hat{V} \hat{c}^\dagger_{\lambda} | G \rangle \ = \ 
   - \mathrm{i} J   \langle G | \hat{a}  \hat{c}_{\mu}^{\dagger } \hat{c}_{\nu}   \hat{c}_\lambda     \tilde{g}_{\nu,\mu} (\hat{a}+ \hat{a}^{\dagger} )  \hat{c}_{\nu}^{\dagger } \hat{c}_{\mu}         \hat{c}^\dagger_{\lambda} | G 
  \rangle \ = \
  - \mathrm{i} J  \tilde{g}_{\nu,\mu} \, .
\end{eqnarray}
The energy difference between the initial and intermediate state is $\epsilon_{\mu} - \epsilon_{\nu} - \hbar \omega_{cav} <0$. For this diagonal process we obtain:
\begin{eqnarray}
\tilde{\Gamma}^{(\mathrm{Type \ II,diag.})}_{\lambda,\lambda} \ &=& \  \sum_{\epsilon_{\nu} > E_F} (1 - \delta_{\lambda,\nu})  \sum_{\epsilon_{\mu} < E_F}  \frac{ J^2\, \tilde{g}_{\nu,\mu} \tilde{g}_{\mu,\nu}}{ \epsilon_{\nu} - \epsilon_{\mu}   + \hbar \omega_{cav}}  \nonumber \\
\ &=& \ \sum_{\epsilon_{\nu} > E_F}  \sum_{\epsilon_{\mu} < E_F}  \frac{ J^2\, \tilde{g}_{\nu,\mu} \tilde{g}_{\mu,\nu}}{ \epsilon_{\nu} - \epsilon_{\mu}   + \hbar \omega_{cav}} 
 \ - \
 \sum_{\epsilon_{\mu} < E_F}  \frac{ J^2\, \tilde{g}_{\lambda,\mu} \tilde{g}_{\mu,\lambda}}{ \epsilon_{\lambda} - \epsilon_{\mu}   + \hbar \omega_{cav}} .
\end{eqnarray}
The first term is independent of $\lambda$ and is therefore a constant diagonal term which can be ignored. The second term exactly cancels the $ \tilde{\Gamma}^{(\mathrm{Type \ II})}_{\lambda,\lambda'}$ term that multiplies $ \delta_{\lambda,\lambda'} $. Finally, we obtain the expression for the total cavity mediated hopping rates:
\begin{equation}
\tilde{\Gamma}_{\lambda,\lambda'} \ \simeq \ \tilde{\Gamma}^{(\mathrm{Type \ I})}_{\lambda,\lambda} +  \tilde{\Gamma}^{(\mathrm{Type \ II})}_{\lambda,\lambda'} +
\delta_{\lambda,\lambda'} \tilde{\Gamma}^{(\mathrm{Type \ II,diag.})}_{\lambda,\lambda} \ = \ 
\sum_{\epsilon_{\mu} \geq E_F}   \frac{  J^2\, \tilde{g}_{\lambda,\mu} \tilde{g}_{\mu,\lambda'}}{ \epsilon_{\mu} - \epsilon_{\lambda}   + \hbar \omega_{cav}} 
- \sum_{\epsilon_{\mu} < E_F}  \frac{ J^2\, \tilde{g}_{\lambda,\mu} \tilde{g}_{\mu,\lambda'}}{ \epsilon_{\lambda'} - \epsilon_{\mu}   + \hbar \omega_{cav}}.
\end{equation}
Given that we are interested in the quantum transport of electrons injected at the Fermi level, the relevant states are such that $\epsilon_{\lambda} \approx E_F$ and  $\epsilon_{\lambda'} \approx E_F$. This allows us to make the following approximations:
\begin{eqnarray}
\tilde{\Gamma}_{\lambda,\lambda'} \ &\simeq& \  
\sum_{\epsilon_{\mu} \geq E_F}   \frac{  J^2\, \tilde{g}_{\lambda,\mu} \tilde{g}_{\mu,\lambda'}}{ |\epsilon_{\mu} - \epsilon_{\lambda}|   + \hbar \omega_{cav}} 
- \sum_{\epsilon_{\mu} < E_F}  \frac{ J^2\, \tilde{g}_{\lambda,\mu} \tilde{g}_{\mu,\lambda'}}{ |\epsilon_{\mu} - \epsilon_{\lambda'} |  + \hbar \omega_{cav}} \nonumber \\
\ &\simeq& \  
\sum_{\epsilon_{\mu} \geq E_F}   \frac{  J^2\, \tilde{g}_{\lambda,\mu} \tilde{g}_{\mu,\lambda'}}{ |\epsilon_\mu - \frac{\epsilon_{\lambda}+\epsilon_{\lambda'}}{2} |   + \hbar \omega_{cav}} 
- \sum_{\epsilon_{\mu} < E_F}  \frac{ J^2\, \tilde{g}_{\lambda,\mu} \tilde{g}_{\mu,\lambda'}}{ |\epsilon_\mu - \frac{\epsilon_{\lambda}+\epsilon_{\lambda'}}{2} |  + \hbar \omega_{cav}} \ = \ J^2 \sum_{\mu}  \frac{\mathrm{sgn}(\epsilon_{\mu}-E_F  )   \, \tilde{g}_{\lambda,\mu} \tilde{g}_{\mu,\lambda'} }{ |\epsilon_\mu - \frac{\epsilon_{\lambda}+\epsilon_{\lambda'}}{2} |  + \hbar \omega_{cav}} , 
\end{eqnarray}
\end{widetext}
where in the final sum $\mu$ labels single-particle states with energy both below and above $E_F$. Finally, we approximate the sign function as $\mathrm{sgn}(\epsilon_{\mu}- E_F) \simeq \mathrm{sgn}(\epsilon_{\mu}- \mathrm{min}(\epsilon_{\lambda},\epsilon_{\lambda'}))$ to obtain Eq. (\ref{gammaTot}).

\section{Conductance for additional 1D potentials}

\begin{figure*}
	\includegraphics[width=1\hsize]{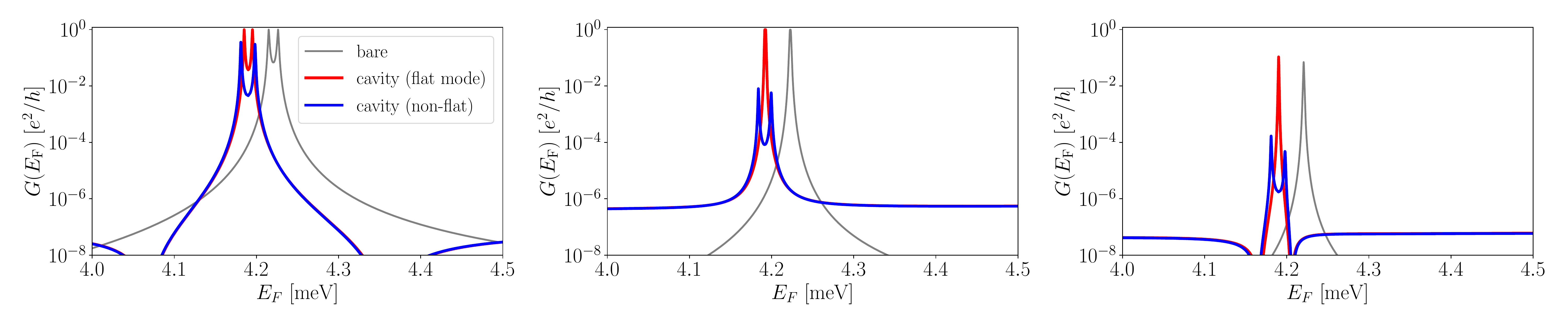}
	\caption{This figure reports the conductance for a 1D symmetric double well potential as in Fig. \ref{double_suppression}. For sake of comparison, the center panel contains the same plot as in Fig. \ref{double_suppression}. Left panel: same, but with a width of the middle potential barrier $25 \%$ thinner than in Fig. \ref{double_suppression}. Right panel: same but with barrier width that is $25 \%$ thicker than in the middle panel.  All other parameters are the same as in Fig. \ref{double_suppression}. \label{fgSM1}}
\end{figure*}

In Fig. \ref{fgSM1}, we report numerical results for the conductance when considering a symmetrical double well 1D potential, as in Fig. \ref{double_suppression}, but with different values of the middle barrier thickness. The left (right) panel shows the conductance when the middle barrier width (see Fig. \ref{double_suppression}) is $25 \% $ thinner (thicker). The center panel is identical to Fig. \ref{double_suppression} of the main text and is reproduced here to facilitate the visual comparison. Note that apart from the middle potential barrier, all other portions of $U(x)$ remain unchanged so that  the total sample size of the 1D system is also varied (by $- 10 \%$ and $+ 10 \%$ respectively). 
For the case of a thinner potential barrier (left panel), the tunnel splitting of the lowest energy electronic states is increased, and the two resonances (corresponding to the symmetric and antisymmetric states of the double well) can be resolved: in this case we find that the effect of the cavity is weaker. This is expected as a larger perturbation is needed to break the hybridization of the states confined in the two quantum wells. For the same reason, the effect of the cavity is instead stronger when the barrier is thicker (right panel). The splitting of the two bare levels is extremely small in this case, hence a smaller perturbation can dehybridize the energy levels of the two wells. Interestingly, in this case the transmission at the center of the resonance is not unity, and a cavity with a flat mode leads to an enhancement.

\begin{figure}
	\includegraphics[width=1\hsize]{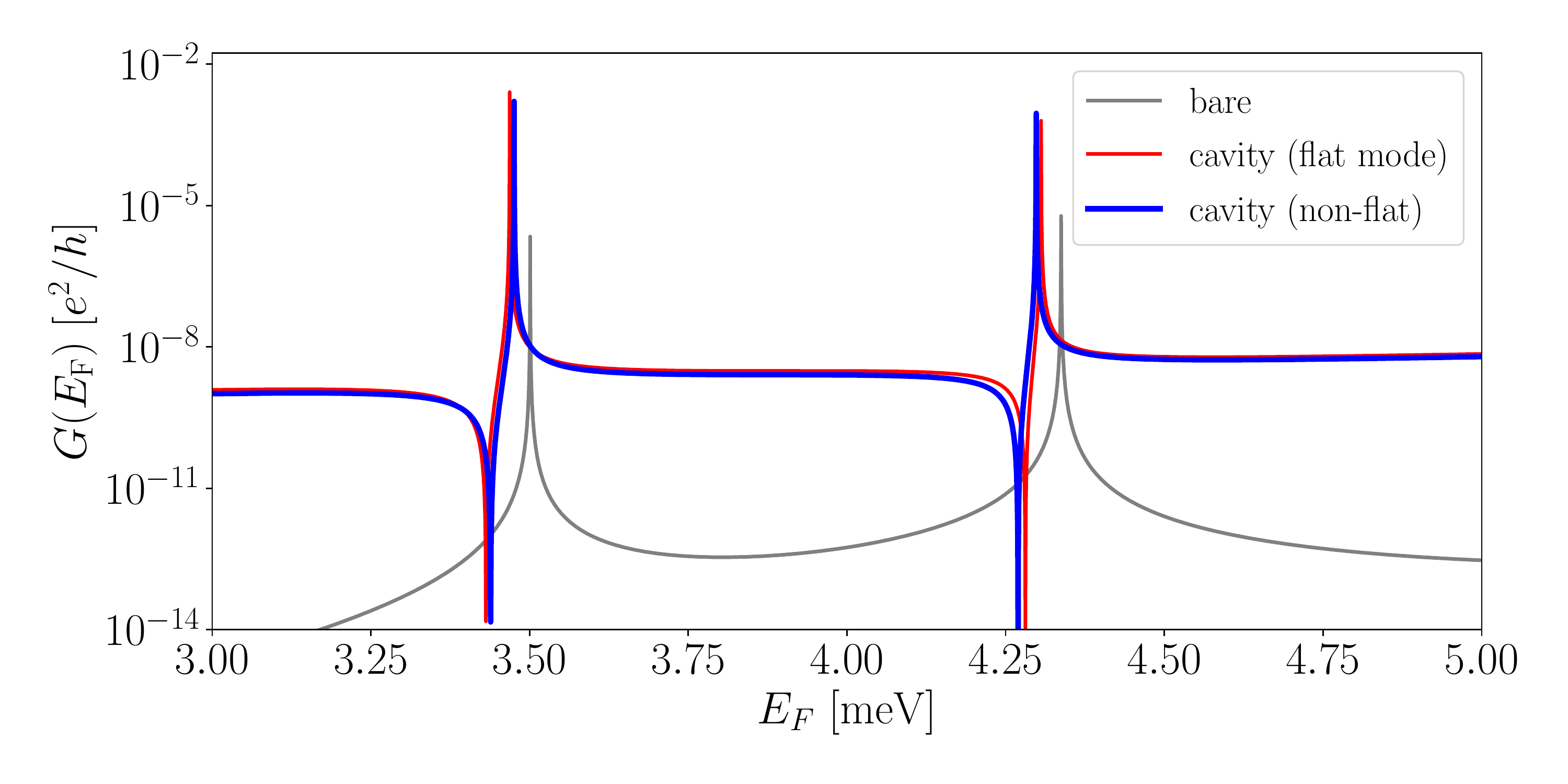}
	\caption{The effect of a cavity field gradient for the asymmetric double well potential case shown in Fig. \ref{double_enhancement}. The blue line shows the conductance for a field with a gradient varying linearly from $0.75$ to $1.25$ times the average value, while the red line shows the conductance for the case of a flat mode. All parameters are the same as in Fig. \ref{double_enhancement}. As show here, in the asymmetric case, the effect of a non-homogeneous cavity field is negligible. \label{fgSM2}}
\end{figure}

In Fig. \ref{fgSM2} we examine the role of the cavity field gradient for the asymmetric potential well considered in Fig. \ref{double_enhancement}. As shown in the figure, the enhanced conductance obtained for a flat  cavity mode is almost unchanged when we consider a cavity field with a gradient (from $0.75$ to $1.25$ times the average value in this example). The key point here is that since the mirror symmetry of the model is already broken by the asymmetric electronic potential, the transmission is less sensitive to the shape of the cavity field. This is in sharp contrast with the case of a symmetric cavity, as shown in Fig. \ref{double_suppression} and Fig. \ref{fgSM1} above, where the same cavity field gradient changes the conductance by orders of magnitude.


\begin{figure*}
	\includegraphics[width=0.325\hsize]{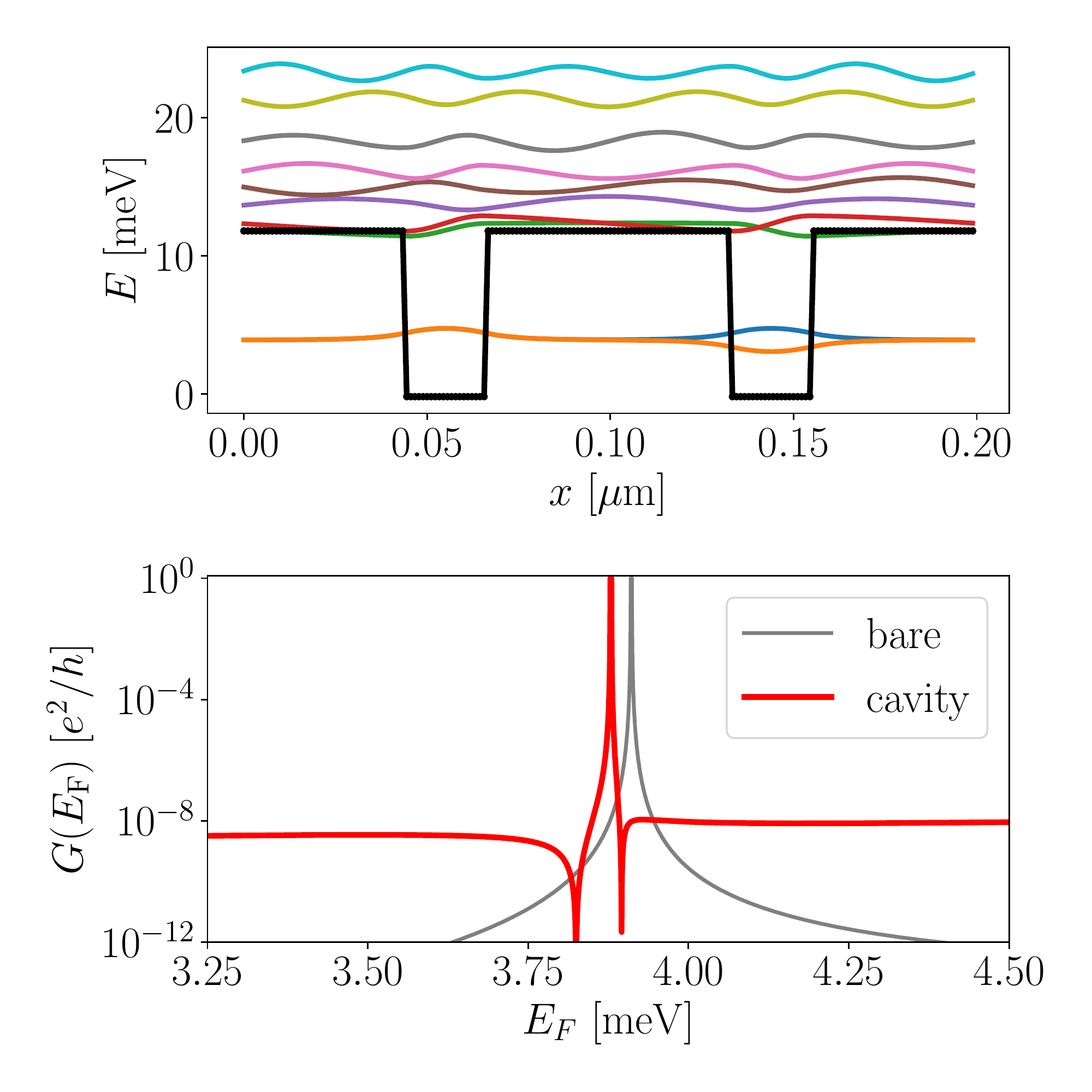}
	\includegraphics[width=0.325\hsize]{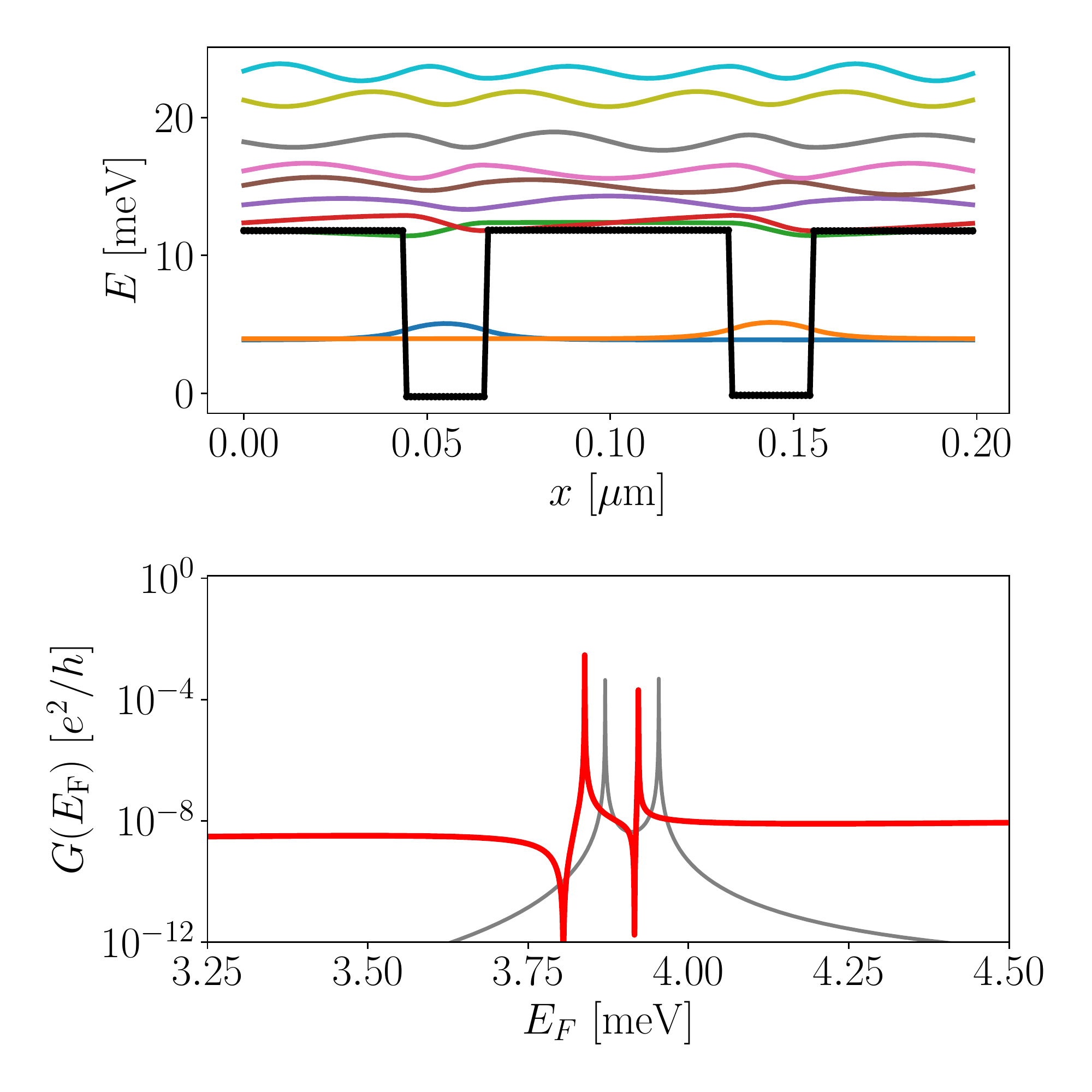}
	\includegraphics[width=0.325\hsize]{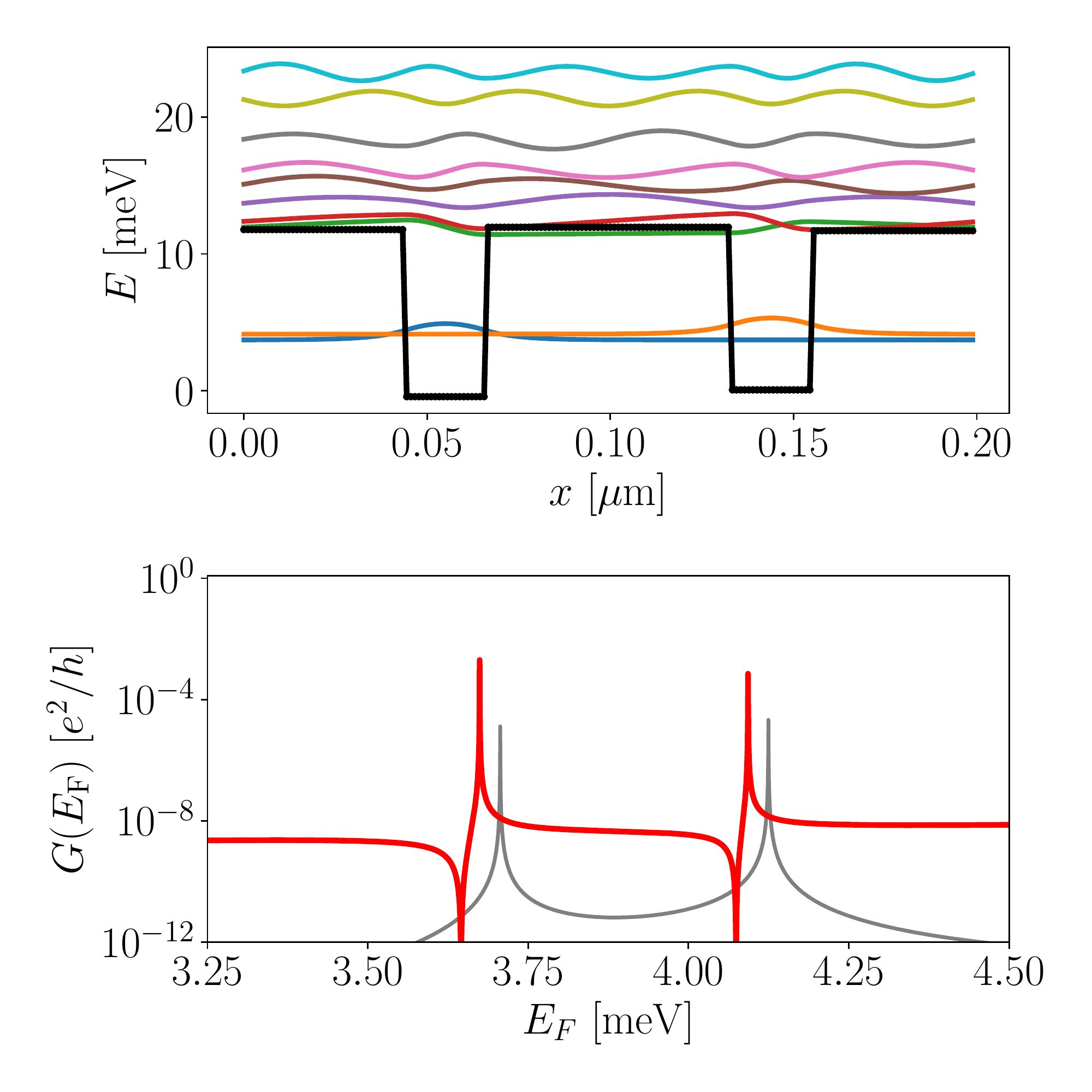}
	\caption{Results showing the effect of asymmetry of the electronic potential. Here we show the conductance for a flat mode and potential $U_s(x)$ with: $s=0 $ (left panel), $ s= 0.1 $ (middle panel) $s= 0.5$ (right panel), where $s$ quantifies the symmetry of the potential (see text). All parameters are the same as in Fig. \ref{double_enhancement}. \label{fgSM3}}
\end{figure*}

In Fig. \ref{fgSM3} we show in more detail the role of asymmetry of the electronic potential. For this end, we write $U_s(x) = U_{S}(x) + s [U_{AS}(x) - U_{S}(x) ]  $, where $U_{AS}(x)$ is the asymmetric potential used in Fig. \ref{double_enhancement}, while $U_{S}(x)$ is a symmetrized version, namely where the left and right wells/barriers are identical. The parameter $s$ quantifies the symmetry, such that $s=0$ corresponds to a perfect symmetry, while $s=1$ corresponds to the asymmetry shown in Fig. \ref{double_enhancement}. In Fig. \ref{fgSM3} we plot the conductance for $s=0, 0.1 , 0.5 $ with a flat cavity mode. Except the potential, all parameters are the same as in Fig. \ref{double_enhancement}. For the perfectly symmetric case (left panel), we find that the cavity changes the lineshape of the resonance, but does not change the peak conductance. This behavior is similar to the flat-mode conductance for the symmetric potentials shown in Fig. \ref{double_suppression} and Fig. \ref{fgSM1}.
For a small asymmetry of $s=0.1$ (middle panel), the bare conductance is strongly suppressed, and the cavity alters the peak conductance. For asymmetry of $s=0.5$ (right panel), the cavity introduces a dramatic enhancement, similar to the one presented in Fig. \ref{double_enhancement}.

\begin{figure*}
	\includegraphics[width=1\hsize]{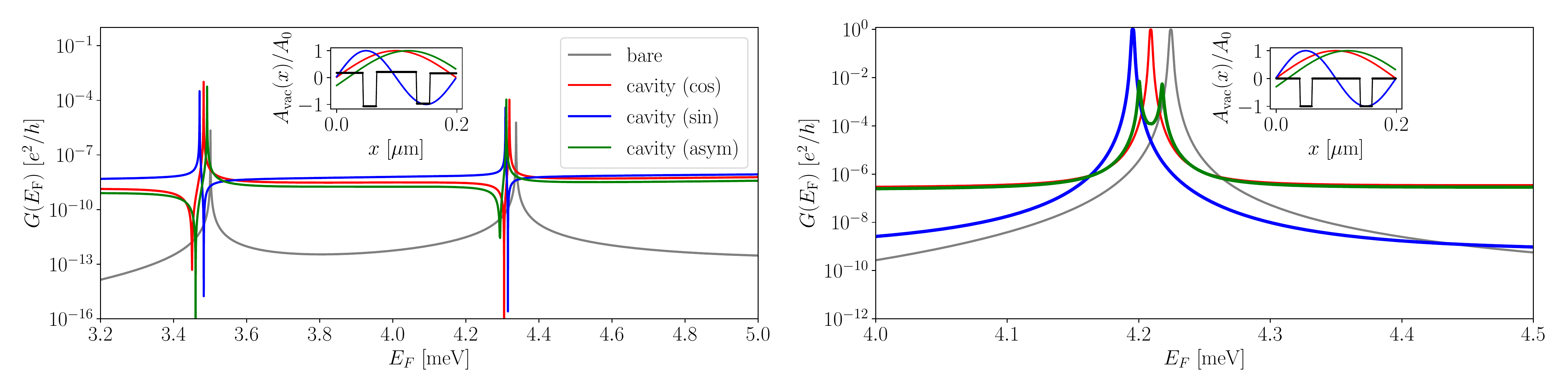}
	\caption{Results for different spatial profiles of the cavity mode. Here we consider cosine (red), sine (blue) and asymmetric (green) cavity modes, as shown in the figure insets. The left panel shows the conductance for the asymmetric potential and the same parameters of Fig. \ref{double_enhancement}. Instead, the right panel shows the conductance for the symmetric potential and the same parameters of Fig. \ref{double_suppression}. \label{fgSM4}}
\end{figure*}

In Fig. \ref{fgSM4} we consider additional cavity field mode shapes (see figure insets). The left and right panels correspond to the electronic potentials of Fig. \ref{double_enhancement} and Fig. \ref{double_suppression} respectively. For the case of the asymmetric potential of Fig. \ref{double_enhancement} (Fig. \ref{fgSM4} left), we find that, qualitatively, the different cavity mode shapes all lead to a similar enhancement. This behavior is similar to the one observed in Fig. \ref{fgSM2}. As we argued above, since the mirror symmetry of the model is already broken by the asymmetric electronic potential, the transmission is less sensitive to the shape of the cavity field. On the contrary, when we consider the symmetric potential of Fig. \ref{double_suppression} (Fig. \ref{fgSM4} right), the transmission is very sensitive to the symmetry of the cavity field. When the cavity field has a cosine/sine shape (red and blue lines), we find that the cavity only shifts the center of the resonance. This is because the cavity fields have the same symmetry as the eigenstates of the bare Hamiltonian, namely symmetric or anti-symmetric with respect to the center. Generally, the matrix element $\tilde{\Gamma}_{\lambda,\lambda'}$ that couples (via the intermediate levels) the symmetric and anti-symmetric states of the double well, will vanish for every symmetric and anti-symmetric cavity field. However, when we consider a cavity field that breaks the symmetry, namely a cosine with a small phase shift (green line) we find that two well states are de-hybridized and the transmission is strongly suppressed. This is similar to the effect of the field gradient in Fig. \ref{double_suppression}.

\section{Details about the conductance calculation}

In this section, we provide additional details about the calculation of the quantum conductance starting from the effective single-particle Hamiltonian.
For a comprehensive review about quantum electron transport, we refer the reader to textbooks \cite{Datta2005,girvin_yang_2019}.

Let us consider a two-terminal configuration, where the system is connected to two semi-infinite leads - source ${\mathcal S}$ and drain ${\mathcal D}$. Our interest is in the linear conductance, that is defined in the limit of vanishing voltage, namely 
\begin{equation}
G= \lim_{V \to 0} I/V \, .
\end{equation}
In this linear and quantum coherent transport regime, the conductance is given by the Landauer-B\"uttiker formula:
\begin{equation}
G(E_{\mathrm{F}}) = \frac{e^2}{h} \sum_{j \in {\mathcal S}\,, j' \in {\mathcal D}}  T_{j,j'}(E_{\mathrm{F}})  \, .    
\end{equation}
Here $T_{j,j'}(E_{\mathrm{F}}) $ is the transmission from one mode (also known as channel) of the source to one mode of the drain. The transmission probabilities are given by the off-diagonal elements of the scattering matrix, $T_{j,j'}(E_{\mathrm{F}})=|S_{j,j'}(E_{\mathrm{F}})|^2$ (for $j \neq j'$).
The scattering matrix can be calculated by a wavefunction approach, namely by solving a scattering problem \cite{Kwant} for an incoming wave from the lead. Alternatively, the scattering matrix can be calculated using Green's function methods. Both approaches are mathematically equivalent due to the Fischer-Lee relation. Generally, we can write the Green's function of our system as:
\begin{eqnarray} 
\mathcal{G} = [ E - \hat{\mathcal H}_{\mathrm{eff}} - \Sigma_{\mathcal S} - \Sigma_{\mathcal D}]^{-1} \, ,
\end{eqnarray}
where $\Sigma_{\mathcal S}$ and $\Sigma_{\mathcal D}$ are the (complex) self-energy terms due to the coupling to the leads. The conductance can be written as:
\begin{equation}
G(E_{\mathrm{F}}) = \frac{e^2}{h} \mathrm{Tr} \left[ \Gamma_{\mathcal S} \mathcal{G}  \Gamma_{\mathcal D} \mathcal{G}^\dagger \right],    
\end{equation}
where $ \Gamma_{\mathcal S}= i ( \Sigma_{\mathcal S} - \Sigma_{\mathcal S}^\dagger )$, and  similarly for $\Gamma_{\mathcal D}$. The matrices $\Sigma_{\mathcal S},\Sigma_{\mathcal D} $ and $\mathcal{G}  $ are all evaluated at $E_F$. For the 1D case, the self-energy terms are given by:
\begin{equation}
\Sigma_{\mathcal S} = -J \vert 1 \rangle \langle 1 \vert  e^{\mathrm{i} k a}  \ \ \  , \ \ \  \Sigma_{\mathcal D} = -J \vert N \rangle \langle N \vert  e^{\mathrm{i} k a} \, , 
\end{equation}
where $ \vert 1 \rangle    $ and $ \vert N \rangle   $ are the leftmost and rightmost sites of the tight-binding grid, where the leads are connected. The wavevector $k$ is determined by the dispersion $  E=-2 J \cos ka  $.
As each lead in 1D has only a single channel, there is only a single transmission coefficient. The conductivity is then given by $G(E_{\mathrm{F}}) = \frac{e^2}{h}   T (E_{\mathrm{F}}) $ with 
\begin{equation}
T (E_{\mathrm{F}})= (2 J \sin k_F a)^2  |  \langle 1 |   \mathcal{G}(E_{\mathrm{F}})   | N  \rangle |^2 .
\end{equation}

For the numerical calculation of the conductance, we used both our code (Figs. \ref{double_enhancement},\ref{double_suppression},\ref{disorder},\ref{fgSM1},\ref{fgSM2},\ref{fgSM3},\ref{fgSM4}), based on the Green's function approach, and the one offered by the package Kwant \cite{Kwant} (Fig. \ref{QH}), based on the wave function approach. We have verified that they give the
same results within the numerical machine precision for 1D.

\bibliography{biblio}

\begin{thebibliography}{45}%
\makeatletter
\providecommand \@ifxundefined [1]{%
 \@ifx{#1\undefined}
}%
\providecommand \@ifnum [1]{%
 \ifnum #1\expandafter \@firstoftwo
 \else \expandafter \@secondoftwo
 \fi
}%
\providecommand \@ifx [1]{%
 \ifx #1\expandafter \@firstoftwo
 \else \expandafter \@secondoftwo
 \fi
}%
\providecommand \natexlab [1]{#1}%
\providecommand \enquote  [1]{``#1''}%
\providecommand \bibnamefont  [1]{#1}%
\providecommand \bibfnamefont [1]{#1}%
\providecommand \citenamefont [1]{#1}%
\providecommand \href@noop [0]{\@secondoftwo}%
\providecommand \href [0]{\begingroup \@sanitize@url \@href}%
\providecommand \@href[1]{\@@startlink{#1}\@@href}%
\providecommand \@@href[1]{\endgroup#1\@@endlink}%
\providecommand \@sanitize@url [0]{\catcode `\\12\catcode `\$12\catcode
  `\&12\catcode `\#12\catcode `\^12\catcode `\_12\catcode `\%12\relax}%
\providecommand \@@startlink[1]{}%
\providecommand \@@endlink[0]{}%
\providecommand \url  [0]{\begingroup\@sanitize@url \@url }%
\providecommand \@url [1]{\endgroup\@href {#1}{\urlprefix }}%
\providecommand \urlprefix  [0]{URL }%
\providecommand \Eprint [0]{\href }%
\providecommand \doibase [0]{https://doi.org/}%
\providecommand \selectlanguage [0]{\@gobble}%
\providecommand \bibinfo  [0]{\@secondoftwo}%
\providecommand \bibfield  [0]{\@secondoftwo}%
\providecommand \translation [1]{[#1]}%
\providecommand \BibitemOpen [0]{}%
\providecommand \bibitemStop [0]{}%
\providecommand \bibitemNoStop [0]{.\EOS\space}%
\providecommand \EOS [0]{\spacefactor3000\relax}%
\providecommand \BibitemShut  [1]{\csname bibitem#1\endcsname}%
\let\auto@bib@innerbib\@empty
\bibitem [{\citenamefont {Genet}\ \emph {et~al.}(2021)\citenamefont {Genet},
  \citenamefont {Faist},\ and\ \citenamefont {Ebbesen}}]{genet2021inducing}%
  \BibitemOpen
  \bibfield  {author} {\bibinfo {author} {\bibfnamefont {C.}~\bibnamefont
  {Genet}}, \bibinfo {author} {\bibfnamefont {J.}~\bibnamefont {Faist}},\ and\
  \bibinfo {author} {\bibfnamefont {T.~W.}\ \bibnamefont {Ebbesen}},\
  }\bibfield  {title} {\bibinfo {title} {Inducing new material properties with
  hybrid light--matter states},\ }\href
  {https://physicstoday.scitation.org/doi/10.1063/PT.3.4749} {\bibfield
  {journal} {\bibinfo  {journal} {Physics Today}\ }\textbf {\bibinfo {volume}
  {74}},\ \bibinfo {pages} {42} (\bibinfo {year} {2021})}\BibitemShut {NoStop}%
\bibitem [{\citenamefont {Garcia-Vidal}\ \emph {et~al.}(2021)\citenamefont
  {Garcia-Vidal}, \citenamefont {Ciuti},\ and\ \citenamefont
  {Ebbesen}}]{garcia2021manipulating}%
  \BibitemOpen
  \bibfield  {author} {\bibinfo {author} {\bibfnamefont {F.~J.}\ \bibnamefont
  {Garcia-Vidal}}, \bibinfo {author} {\bibfnamefont {C.}~\bibnamefont
  {Ciuti}},\ and\ \bibinfo {author} {\bibfnamefont {T.~W.}\ \bibnamefont
  {Ebbesen}},\ }\bibfield  {title} {\bibinfo {title} {Manipulating matter by
  strong coupling to vacuum fields},\ }\href
  {https://doi.org/10.1126/science.abd0336} {\bibfield  {journal} {\bibinfo
  {journal} {Science}\ }\textbf {\bibinfo {volume} {373}},\ \bibinfo {pages}
  {eabd0336} (\bibinfo {year} {2021})}\BibitemShut {NoStop}%
\bibitem [{\citenamefont {Bloch}\ \emph {et~al.}(2022)\citenamefont {Bloch},
  \citenamefont {Cavalleri}, \citenamefont {Galitski}, \citenamefont {Hafezi},\
  and\ \citenamefont {Rubio}}]{bloch2022strongly}%
  \BibitemOpen
  \bibfield  {author} {\bibinfo {author} {\bibfnamefont {J.}~\bibnamefont
  {Bloch}}, \bibinfo {author} {\bibfnamefont {A.}~\bibnamefont {Cavalleri}},
  \bibinfo {author} {\bibfnamefont {V.}~\bibnamefont {Galitski}}, \bibinfo
  {author} {\bibfnamefont {M.}~\bibnamefont {Hafezi}},\ and\ \bibinfo {author}
  {\bibfnamefont {A.}~\bibnamefont {Rubio}},\ }\bibfield  {title} {\bibinfo
  {title} {Strongly correlated electron--photon systems},\ }\href
  {https://www.nature.com/articles/s41586-022-04726-w} {\bibfield  {journal}
  {\bibinfo  {journal} {Nature}\ ,\ \bibinfo {pages} {1}} (\bibinfo {year}
  {2022})}\BibitemShut {NoStop}%
\bibitem [{\citenamefont {Schlawin}\ \emph {et~al.}(2022)\citenamefont
  {Schlawin}, \citenamefont {Kennes},\ and\ \citenamefont
  {Sentef}}]{Schlawin2022}%
  \BibitemOpen
  \bibfield  {author} {\bibinfo {author} {\bibfnamefont {F.}~\bibnamefont
  {Schlawin}}, \bibinfo {author} {\bibfnamefont {D.~M.}\ \bibnamefont
  {Kennes}},\ and\ \bibinfo {author} {\bibfnamefont {M.~A.}\ \bibnamefont
  {Sentef}},\ }\bibfield  {title} {\bibinfo {title} {Cavity quantum
  materials},\ }\href {https://doi.org/10.1063/5.0083825} {\bibfield  {journal}
  {\bibinfo  {journal} {Applied Physics Reviews}\ }\textbf {\bibinfo {volume}
  {9}},\ \bibinfo {pages} {011312} (\bibinfo {year} {2022})}\BibitemShut
  {NoStop}%
\bibitem [{\citenamefont {Nataf}\ \emph {et~al.}(2019)\citenamefont {Nataf},
  \citenamefont {Champel}, \citenamefont {Blatter},\ and\ \citenamefont
  {Basko}}]{Nataf2019}%
  \BibitemOpen
  \bibfield  {author} {\bibinfo {author} {\bibfnamefont {P.}~\bibnamefont
  {Nataf}}, \bibinfo {author} {\bibfnamefont {T.}~\bibnamefont {Champel}},
  \bibinfo {author} {\bibfnamefont {G.}~\bibnamefont {Blatter}},\ and\ \bibinfo
  {author} {\bibfnamefont {D.~M.}\ \bibnamefont {Basko}},\ }\bibfield  {title}
  {\bibinfo {title} {Rashba cavity {QED}: A route towards the superradiant
  quantum phase transition},\ }\href
  {https://doi.org/10.1103/PhysRevLett.123.207402} {\bibfield  {journal}
  {\bibinfo  {journal} {Phys. Rev. Lett.}\ }\textbf {\bibinfo {volume} {123}},\
  \bibinfo {pages} {207402} (\bibinfo {year} {2019})}\BibitemShut {NoStop}%
\bibitem [{\citenamefont {Schlawin}\ \emph {et~al.}(2019)\citenamefont
  {Schlawin}, \citenamefont {Cavalleri},\ and\ \citenamefont
  {Jaksch}}]{Schlawin2019}%
  \BibitemOpen
  \bibfield  {author} {\bibinfo {author} {\bibfnamefont {F.}~\bibnamefont
  {Schlawin}}, \bibinfo {author} {\bibfnamefont {A.}~\bibnamefont
  {Cavalleri}},\ and\ \bibinfo {author} {\bibfnamefont {D.}~\bibnamefont
  {Jaksch}},\ }\bibfield  {title} {\bibinfo {title} {Cavity-mediated
  electron-photon superconductivity},\ }\href
  {https://doi.org/10.1103/PhysRevLett.122.133602} {\bibfield  {journal}
  {\bibinfo  {journal} {Phys. Rev. Lett.}\ }\textbf {\bibinfo {volume} {122}},\
  \bibinfo {pages} {133602} (\bibinfo {year} {2019})}\BibitemShut {NoStop}%
\bibitem [{\citenamefont {Li}\ and\ \citenamefont
  {Eckstein}(2020)}]{Eckstein2020}%
  \BibitemOpen
  \bibfield  {author} {\bibinfo {author} {\bibfnamefont {J.}~\bibnamefont
  {Li}}\ and\ \bibinfo {author} {\bibfnamefont {M.}~\bibnamefont {Eckstein}},\
  }\bibfield  {title} {\bibinfo {title} {Manipulating intertwined orders in
  solids with quantum light},\ }\href
  {https://doi.org/10.1103/PhysRevLett.125.217402} {\bibfield  {journal}
  {\bibinfo  {journal} {Phys. Rev. Lett.}\ }\textbf {\bibinfo {volume} {125}},\
  \bibinfo {pages} {217402} (\bibinfo {year} {2020})}\BibitemShut {NoStop}%
\bibitem [{\citenamefont {Guerci}\ \emph {et~al.}(2020)\citenamefont {Guerci},
  \citenamefont {Simon},\ and\ \citenamefont {Mora}}]{Guerci2020}%
  \BibitemOpen
  \bibfield  {author} {\bibinfo {author} {\bibfnamefont {D.}~\bibnamefont
  {Guerci}}, \bibinfo {author} {\bibfnamefont {P.}~\bibnamefont {Simon}},\ and\
  \bibinfo {author} {\bibfnamefont {C.}~\bibnamefont {Mora}},\ }\bibfield
  {title} {\bibinfo {title} {Superradiant phase transition in electronic
  systems and emergent topological phases},\ }\href
  {https://doi.org/10.1103/PhysRevLett.125.257604} {\bibfield  {journal}
  {\bibinfo  {journal} {Phys. Rev. Lett.}\ }\textbf {\bibinfo {volume} {125}},\
  \bibinfo {pages} {257604} (\bibinfo {year} {2020})}\BibitemShut {NoStop}%
\bibitem [{\citenamefont {Andolina}\ \emph {et~al.}(2020)\citenamefont
  {Andolina}, \citenamefont {Pellegrino}, \citenamefont {Giovannetti},
  \citenamefont {MacDonald},\ and\ \citenamefont {Polini}}]{Andolina2020}%
  \BibitemOpen
  \bibfield  {author} {\bibinfo {author} {\bibfnamefont {G.~M.}\ \bibnamefont
  {Andolina}}, \bibinfo {author} {\bibfnamefont {F.~M.~D.}\ \bibnamefont
  {Pellegrino}}, \bibinfo {author} {\bibfnamefont {V.}~\bibnamefont
  {Giovannetti}}, \bibinfo {author} {\bibfnamefont {A.~H.}\ \bibnamefont
  {MacDonald}},\ and\ \bibinfo {author} {\bibfnamefont {M.}~\bibnamefont
  {Polini}},\ }\bibfield  {title} {\bibinfo {title} {Theory of photon
  condensation in a spatially varying electromagnetic field},\ }\href
  {https://doi.org/10.1103/PhysRevB.102.125137} {\bibfield  {journal} {\bibinfo
   {journal} {Phys. Rev. B}\ }\textbf {\bibinfo {volume} {102}},\ \bibinfo
  {pages} {125137} (\bibinfo {year} {2020})}\BibitemShut {NoStop}%
\bibitem [{\citenamefont {Amelio}\ \emph {et~al.}(2021)\citenamefont {Amelio},
  \citenamefont {Korosec}, \citenamefont {Carusotto},\ and\ \citenamefont
  {Mazza}}]{Amelio2021}%
  \BibitemOpen
  \bibfield  {author} {\bibinfo {author} {\bibfnamefont {I.}~\bibnamefont
  {Amelio}}, \bibinfo {author} {\bibfnamefont {L.}~\bibnamefont {Korosec}},
  \bibinfo {author} {\bibfnamefont {I.}~\bibnamefont {Carusotto}},\ and\
  \bibinfo {author} {\bibfnamefont {G.}~\bibnamefont {Mazza}},\ }\bibfield
  {title} {\bibinfo {title} {Optical dressing of the electronic response of
  two-dimensional semiconductors in quantum and classical descriptions of
  cavity electrodynamics},\ }\href
  {https://doi.org/10.1103/PhysRevB.104.235120} {\bibfield  {journal} {\bibinfo
   {journal} {Phys. Rev. B}\ }\textbf {\bibinfo {volume} {104}},\ \bibinfo
  {pages} {235120} (\bibinfo {year} {2021})}\BibitemShut {NoStop}%
\bibitem [{\citenamefont {Orgiu}\ \emph {et~al.}(2015)\citenamefont {Orgiu},
  \citenamefont {George}, \citenamefont {Hutchison}, \citenamefont {Devaux},
  \citenamefont {Dayen}, \citenamefont {Doudin}, \citenamefont {Stellacci},
  \citenamefont {Genet}, \citenamefont {Schachenmayer}, \citenamefont {Genes},
  \citenamefont {Pupillo}, \citenamefont {Samor{\`{\i}}},\ and\ \citenamefont
  {Ebbesen}}]{orgiu2015conductivity}%
  \BibitemOpen
  \bibfield  {author} {\bibinfo {author} {\bibfnamefont {E.}~\bibnamefont
  {Orgiu}}, \bibinfo {author} {\bibfnamefont {J.}~\bibnamefont {George}},
  \bibinfo {author} {\bibfnamefont {J.~A.}\ \bibnamefont {Hutchison}}, \bibinfo
  {author} {\bibfnamefont {E.}~\bibnamefont {Devaux}}, \bibinfo {author}
  {\bibfnamefont {J.~F.}\ \bibnamefont {Dayen}}, \bibinfo {author}
  {\bibfnamefont {B.}~\bibnamefont {Doudin}}, \bibinfo {author} {\bibfnamefont
  {F.}~\bibnamefont {Stellacci}}, \bibinfo {author} {\bibfnamefont
  {C.}~\bibnamefont {Genet}}, \bibinfo {author} {\bibfnamefont
  {J.}~\bibnamefont {Schachenmayer}}, \bibinfo {author} {\bibfnamefont
  {C.}~\bibnamefont {Genes}}, \bibinfo {author} {\bibfnamefont
  {G.}~\bibnamefont {Pupillo}}, \bibinfo {author} {\bibfnamefont
  {P.}~\bibnamefont {Samor{\`{\i}}}},\ and\ \bibinfo {author} {\bibfnamefont
  {T.~W.}\ \bibnamefont {Ebbesen}},\ }\bibfield  {title} {\bibinfo {title}
  {Conductivity in organic semiconductors hybridized with the vacuum field},\
  }\href {https://doi.org/10.1038/nmat4392} {\bibfield  {journal} {\bibinfo
  {journal} {Nature Materials}\ }\textbf {\bibinfo {volume} {14}},\ \bibinfo
  {pages} {1123} (\bibinfo {year} {2015})}\BibitemShut {NoStop}%
\bibitem [{\citenamefont {Nagarajan}\ \emph {et~al.}(2020)\citenamefont
  {Nagarajan}, \citenamefont {George}, \citenamefont {Thomas}, \citenamefont
  {Devaux}, \citenamefont {Chervy}, \citenamefont {Azzini}, \citenamefont
  {Joseph}, \citenamefont {Jouaiti}, \citenamefont {Hosseini}, \citenamefont
  {Kumar}, \citenamefont {Genet}, \citenamefont {Bartolo}, \citenamefont
  {Ciuti},\ and\ \citenamefont {Ebbesen}}]{nagarajan2020conductivity}%
  \BibitemOpen
  \bibfield  {author} {\bibinfo {author} {\bibfnamefont {K.}~\bibnamefont
  {Nagarajan}}, \bibinfo {author} {\bibfnamefont {J.}~\bibnamefont {George}},
  \bibinfo {author} {\bibfnamefont {A.}~\bibnamefont {Thomas}}, \bibinfo
  {author} {\bibfnamefont {E.}~\bibnamefont {Devaux}}, \bibinfo {author}
  {\bibfnamefont {T.}~\bibnamefont {Chervy}}, \bibinfo {author} {\bibfnamefont
  {S.}~\bibnamefont {Azzini}}, \bibinfo {author} {\bibfnamefont
  {K.}~\bibnamefont {Joseph}}, \bibinfo {author} {\bibfnamefont
  {A.}~\bibnamefont {Jouaiti}}, \bibinfo {author} {\bibfnamefont {M.~W.}\
  \bibnamefont {Hosseini}}, \bibinfo {author} {\bibfnamefont {A.}~\bibnamefont
  {Kumar}}, \bibinfo {author} {\bibfnamefont {C.}~\bibnamefont {Genet}},
  \bibinfo {author} {\bibfnamefont {N.}~\bibnamefont {Bartolo}}, \bibinfo
  {author} {\bibfnamefont {C.}~\bibnamefont {Ciuti}},\ and\ \bibinfo {author}
  {\bibfnamefont {T.~W.}\ \bibnamefont {Ebbesen}},\ }\bibfield  {title}
  {\bibinfo {title} {Conductivity and photoconductivity of a p-type organic
  semiconductor under ultrastrong coupling},\ }\href
  {https://doi.org/10.1021/acsnano.0c03496} {\bibfield  {journal} {\bibinfo
  {journal} {{ACS} Nano}\ }\textbf {\bibinfo {volume} {14}},\ \bibinfo {pages}
  {10219} (\bibinfo {year} {2020})}\BibitemShut {NoStop}%
\bibitem [{\citenamefont {Paravicini-Bagliani}\ \emph
  {et~al.}(2018)\citenamefont {Paravicini-Bagliani}, \citenamefont
  {Appugliese}, \citenamefont {Richter}, \citenamefont {Valmorra},
  \citenamefont {Keller}, \citenamefont {Beck}, \citenamefont {Bartolo},
  \citenamefont {R\"{o}ssler}, \citenamefont {Ihn}, \citenamefont {Ensslin},
  \citenamefont {Ciuti}, \citenamefont {Scalari},\ and\ \citenamefont
  {Faist}}]{ParaviciniBagliani2018}%
  \BibitemOpen
  \bibfield  {author} {\bibinfo {author} {\bibfnamefont {G.~L.}\ \bibnamefont
  {Paravicini-Bagliani}}, \bibinfo {author} {\bibfnamefont {F.}~\bibnamefont
  {Appugliese}}, \bibinfo {author} {\bibfnamefont {E.}~\bibnamefont {Richter}},
  \bibinfo {author} {\bibfnamefont {F.}~\bibnamefont {Valmorra}}, \bibinfo
  {author} {\bibfnamefont {J.}~\bibnamefont {Keller}}, \bibinfo {author}
  {\bibfnamefont {M.}~\bibnamefont {Beck}}, \bibinfo {author} {\bibfnamefont
  {N.}~\bibnamefont {Bartolo}}, \bibinfo {author} {\bibfnamefont
  {C.}~\bibnamefont {R\"{o}ssler}}, \bibinfo {author} {\bibfnamefont
  {T.}~\bibnamefont {Ihn}}, \bibinfo {author} {\bibfnamefont {K.}~\bibnamefont
  {Ensslin}}, \bibinfo {author} {\bibfnamefont {C.}~\bibnamefont {Ciuti}},
  \bibinfo {author} {\bibfnamefont {G.}~\bibnamefont {Scalari}},\ and\ \bibinfo
  {author} {\bibfnamefont {J.}~\bibnamefont {Faist}},\ }\bibfield  {title}
  {\bibinfo {title} {Magneto-transport controlled by {L}andau polariton
  states},\ }\href {https://doi.org/10.1038/s41567-018-0346-y} {\bibfield
  {journal} {\bibinfo  {journal} {Nature Physics}\ }\textbf {\bibinfo {volume}
  {15}},\ \bibinfo {pages} {186} (\bibinfo {year} {2018})}\BibitemShut
  {NoStop}%
\bibitem [{\citenamefont {Appugliese}\ \emph {et~al.}(2022)\citenamefont
  {Appugliese}, \citenamefont {Enkner}, \citenamefont {Paravicini-Bagliani},
  \citenamefont {Beck}, \citenamefont {Reichl}, \citenamefont {Wegscheider},
  \citenamefont {Scalari}, \citenamefont {Ciuti},\ and\ \citenamefont
  {Faist}}]{Appugliese2022}%
  \BibitemOpen
  \bibfield  {author} {\bibinfo {author} {\bibfnamefont {F.}~\bibnamefont
  {Appugliese}}, \bibinfo {author} {\bibfnamefont {J.}~\bibnamefont {Enkner}},
  \bibinfo {author} {\bibfnamefont {G.~L.}\ \bibnamefont
  {Paravicini-Bagliani}}, \bibinfo {author} {\bibfnamefont {M.}~\bibnamefont
  {Beck}}, \bibinfo {author} {\bibfnamefont {C.}~\bibnamefont {Reichl}},
  \bibinfo {author} {\bibfnamefont {W.}~\bibnamefont {Wegscheider}}, \bibinfo
  {author} {\bibfnamefont {G.}~\bibnamefont {Scalari}}, \bibinfo {author}
  {\bibfnamefont {C.}~\bibnamefont {Ciuti}},\ and\ \bibinfo {author}
  {\bibfnamefont {J.}~\bibnamefont {Faist}},\ }\bibfield  {title} {\bibinfo
  {title} {Breakdown of topological protection by cavity vacuum fields in the
  integer quantum {H}all effect},\ }\href
  {https://doi.org/10.1126/science.abl5818} {\bibfield  {journal} {\bibinfo
  {journal} {Science}\ }\textbf {\bibinfo {volume} {375}},\ \bibinfo {pages}
  {1030} (\bibinfo {year} {2022})}\BibitemShut {NoStop}%
\bibitem [{\citenamefont {Valmorra}\ \emph {et~al.}(2021)\citenamefont
  {Valmorra}, \citenamefont {Yoshida}, \citenamefont {Contamin}, \citenamefont
  {Messelot}, \citenamefont {Massabeau}, \citenamefont {Delbecq}, \citenamefont
  {Dartiailh}, \citenamefont {Desjardins}, \citenamefont {Cubaynes},
  \citenamefont {Leghtas}, \citenamefont {Hirakawa}, \citenamefont {Tignon},
  \citenamefont {Dhillon}, \citenamefont {Balibar}, \citenamefont {Mangeney},
  \citenamefont {Cottet},\ and\ \citenamefont {Kontos}}]{Valmorra2021}%
  \BibitemOpen
  \bibfield  {author} {\bibinfo {author} {\bibfnamefont {F.}~\bibnamefont
  {Valmorra}}, \bibinfo {author} {\bibfnamefont {K.}~\bibnamefont {Yoshida}},
  \bibinfo {author} {\bibfnamefont {L.~C.}\ \bibnamefont {Contamin}}, \bibinfo
  {author} {\bibfnamefont {S.}~\bibnamefont {Messelot}}, \bibinfo {author}
  {\bibfnamefont {S.}~\bibnamefont {Massabeau}}, \bibinfo {author}
  {\bibfnamefont {M.~R.}\ \bibnamefont {Delbecq}}, \bibinfo {author}
  {\bibfnamefont {M.~C.}\ \bibnamefont {Dartiailh}}, \bibinfo {author}
  {\bibfnamefont {M.~M.}\ \bibnamefont {Desjardins}}, \bibinfo {author}
  {\bibfnamefont {T.}~\bibnamefont {Cubaynes}}, \bibinfo {author}
  {\bibfnamefont {Z.}~\bibnamefont {Leghtas}}, \bibinfo {author} {\bibfnamefont
  {K.}~\bibnamefont {Hirakawa}}, \bibinfo {author} {\bibfnamefont
  {J.}~\bibnamefont {Tignon}}, \bibinfo {author} {\bibfnamefont
  {S.}~\bibnamefont {Dhillon}}, \bibinfo {author} {\bibfnamefont
  {S.}~\bibnamefont {Balibar}}, \bibinfo {author} {\bibfnamefont
  {J.}~\bibnamefont {Mangeney}}, \bibinfo {author} {\bibfnamefont
  {A.}~\bibnamefont {Cottet}},\ and\ \bibinfo {author} {\bibfnamefont
  {T.}~\bibnamefont {Kontos}},\ }\bibfield  {title} {\bibinfo {title}
  {Vacuum-field-induced {THz} transport gap in a carbon nanotube quantum dot},\
  }\bibfield  {journal} {\bibinfo  {journal} {Nature Communications}\ }\textbf
  {\bibinfo {volume} {12}},\ \href {https://doi.org/10.1038/s41467-021-25733-x}
  {10.1038/s41467-021-25733-x} (\bibinfo {year} {2021})\BibitemShut {NoStop}%
\bibitem [{\citenamefont {Hagenm\"uller}\ \emph {et~al.}(2017)\citenamefont
  {Hagenm\"uller}, \citenamefont {Schachenmayer}, \citenamefont {Sch\"utz},
  \citenamefont {Genes},\ and\ \citenamefont
  {Pupillo}}]{hagenmuller2017cavity}%
  \BibitemOpen
  \bibfield  {author} {\bibinfo {author} {\bibfnamefont {D.}~\bibnamefont
  {Hagenm\"uller}}, \bibinfo {author} {\bibfnamefont {J.}~\bibnamefont
  {Schachenmayer}}, \bibinfo {author} {\bibfnamefont {S.}~\bibnamefont
  {Sch\"utz}}, \bibinfo {author} {\bibfnamefont {C.}~\bibnamefont {Genes}},\
  and\ \bibinfo {author} {\bibfnamefont {G.}~\bibnamefont {Pupillo}},\
  }\bibfield  {title} {\bibinfo {title} {Cavity-enhanced transport of charge},\
  }\href {https://doi.org/10.1103/PhysRevLett.119.223601} {\bibfield  {journal}
  {\bibinfo  {journal} {Phys. Rev. Lett.}\ }\textbf {\bibinfo {volume} {119}},\
  \bibinfo {pages} {223601} (\bibinfo {year} {2017})}\BibitemShut {NoStop}%
\bibitem [{\citenamefont {Hagenm\"uller}\ \emph {et~al.}(2018)\citenamefont
  {Hagenm\"uller}, \citenamefont {Sch\"utz}, \citenamefont {Schachenmayer},
  \citenamefont {Genes},\ and\ \citenamefont {Pupillo}}]{Hagen_PRB_2018}%
  \BibitemOpen
  \bibfield  {author} {\bibinfo {author} {\bibfnamefont {D.}~\bibnamefont
  {Hagenm\"uller}}, \bibinfo {author} {\bibfnamefont {S.}~\bibnamefont
  {Sch\"utz}}, \bibinfo {author} {\bibfnamefont {J.}~\bibnamefont
  {Schachenmayer}}, \bibinfo {author} {\bibfnamefont {C.}~\bibnamefont
  {Genes}},\ and\ \bibinfo {author} {\bibfnamefont {G.}~\bibnamefont
  {Pupillo}},\ }\bibfield  {title} {\bibinfo {title} {Cavity-assisted
  mesoscopic transport of fermions: Coherent and dissipative dynamics},\ }\href
  {https://doi.org/10.1103/PhysRevB.97.205303} {\bibfield  {journal} {\bibinfo
  {journal} {Phys. Rev. B}\ }\textbf {\bibinfo {volume} {97}},\ \bibinfo
  {pages} {205303} (\bibinfo {year} {2018})}\BibitemShut {NoStop}%
\bibitem [{\citenamefont {Ch\'avez}\ \emph {et~al.}(2021)\citenamefont
  {Ch\'avez}, \citenamefont {Mattiotti}, \citenamefont {M\'endez-Berm\'udez},
  \citenamefont {Borgonovi},\ and\ \citenamefont {Celardo}}]{Chavez2021}%
  \BibitemOpen
  \bibfield  {author} {\bibinfo {author} {\bibfnamefont {N.~C.}\ \bibnamefont
  {Ch\'avez}}, \bibinfo {author} {\bibfnamefont {F.}~\bibnamefont {Mattiotti}},
  \bibinfo {author} {\bibfnamefont {J.~A.}\ \bibnamefont
  {M\'endez-Berm\'udez}}, \bibinfo {author} {\bibfnamefont {F.}~\bibnamefont
  {Borgonovi}},\ and\ \bibinfo {author} {\bibfnamefont {G.~L.}\ \bibnamefont
  {Celardo}},\ }\bibfield  {title} {\bibinfo {title} {Disorder-enhanced and
  disorder-independent transport with long-range hopping: Application to
  molecular chains in optical cavities},\ }\href
  {https://doi.org/10.1103/PhysRevLett.126.153201} {\bibfield  {journal}
  {\bibinfo  {journal} {Phys. Rev. Lett.}\ }\textbf {\bibinfo {volume} {126}},\
  \bibinfo {pages} {153201} (\bibinfo {year} {2021})}\BibitemShut {NoStop}%
\bibitem [{\citenamefont {De~Liberato}\ and\ \citenamefont
  {Ciuti}(2008)}]{DeLib2008}%
  \BibitemOpen
  \bibfield  {author} {\bibinfo {author} {\bibfnamefont {S.}~\bibnamefont
  {De~Liberato}}\ and\ \bibinfo {author} {\bibfnamefont {C.}~\bibnamefont
  {Ciuti}},\ }\bibfield  {title} {\bibinfo {title} {Quantum model of
  microcavity intersubband electroluminescent devices},\ }\href
  {https://doi.org/10.1103/PhysRevB.77.155321} {\bibfield  {journal} {\bibinfo
  {journal} {Phys. Rev. B}\ }\textbf {\bibinfo {volume} {77}},\ \bibinfo
  {pages} {155321} (\bibinfo {year} {2008})}\BibitemShut {NoStop}%
\bibitem [{\citenamefont {De~Liberato}\ and\ \citenamefont
  {Ciuti}(2009)}]{DeLib2009}%
  \BibitemOpen
  \bibfield  {author} {\bibinfo {author} {\bibfnamefont {S.}~\bibnamefont
  {De~Liberato}}\ and\ \bibinfo {author} {\bibfnamefont {C.}~\bibnamefont
  {Ciuti}},\ }\bibfield  {title} {\bibinfo {title} {Quantum theory of electron
  tunneling into intersubband cavity polariton states},\ }\href
  {https://doi.org/10.1103/PhysRevB.79.075317} {\bibfield  {journal} {\bibinfo
  {journal} {Phys. Rev. B}\ }\textbf {\bibinfo {volume} {79}},\ \bibinfo
  {pages} {075317} (\bibinfo {year} {2009})}\BibitemShut {NoStop}%
\bibitem [{\citenamefont {Bartolo}\ and\ \citenamefont
  {Ciuti}(2018)}]{Bartolo_2018}%
  \BibitemOpen
  \bibfield  {author} {\bibinfo {author} {\bibfnamefont {N.}~\bibnamefont
  {Bartolo}}\ and\ \bibinfo {author} {\bibfnamefont {C.}~\bibnamefont
  {Ciuti}},\ }\bibfield  {title} {\bibinfo {title} {Vacuum-dressed cavity
  magnetotransport of a two-dimensional electron gas},\ }\href
  {https://doi.org/10.1103/PhysRevB.98.205301} {\bibfield  {journal} {\bibinfo
  {journal} {Phys. Rev. B}\ }\textbf {\bibinfo {volume} {98}},\ \bibinfo
  {pages} {205301} (\bibinfo {year} {2018})}\BibitemShut {NoStop}%
\bibitem [{\citenamefont {Naudet-Baulieu}\ \emph {et~al.}(2019)\citenamefont
  {Naudet-Baulieu}, \citenamefont {Bartolo}, \citenamefont {Orso},\ and\
  \citenamefont {Ciuti}}]{NaudetBaulieu2019}%
  \BibitemOpen
  \bibfield  {author} {\bibinfo {author} {\bibfnamefont {C.}~\bibnamefont
  {Naudet-Baulieu}}, \bibinfo {author} {\bibfnamefont {N.}~\bibnamefont
  {Bartolo}}, \bibinfo {author} {\bibfnamefont {G.}~\bibnamefont {Orso}},\ and\
  \bibinfo {author} {\bibfnamefont {C.}~\bibnamefont {Ciuti}},\ }\bibfield
  {title} {\bibinfo {title} {Dark vertical conductance of cavity-embedded
  semiconductor heterostructures},\ }\href
  {https://doi.org/10.1088/1367-2630/ab41c2} {\bibfield  {journal} {\bibinfo
  {journal} {New Journal of Physics}\ }\textbf {\bibinfo {volume} {21}},\
  \bibinfo {pages} {093061} (\bibinfo {year} {2019})}\BibitemShut {NoStop}%
\bibitem [{\citenamefont {Paulillo}\ \emph {et~al.}(2017)\citenamefont
  {Paulillo}, \citenamefont {Pirotta}, \citenamefont {Nong}, \citenamefont
  {Crozat}, \citenamefont {Guilet}, \citenamefont {Xu}, \citenamefont
  {Dhillon}, \citenamefont {Li}, \citenamefont {Davies}, \citenamefont
  {Linfield},\ and\ \citenamefont {Colombelli}}]{Paulillo:17}%
  \BibitemOpen
  \bibfield  {author} {\bibinfo {author} {\bibfnamefont {B.}~\bibnamefont
  {Paulillo}}, \bibinfo {author} {\bibfnamefont {S.}~\bibnamefont {Pirotta}},
  \bibinfo {author} {\bibfnamefont {H.}~\bibnamefont {Nong}}, \bibinfo {author}
  {\bibfnamefont {P.}~\bibnamefont {Crozat}}, \bibinfo {author} {\bibfnamefont
  {S.}~\bibnamefont {Guilet}}, \bibinfo {author} {\bibfnamefont
  {G.}~\bibnamefont {Xu}}, \bibinfo {author} {\bibfnamefont {S.}~\bibnamefont
  {Dhillon}}, \bibinfo {author} {\bibfnamefont {L.~H.}\ \bibnamefont {Li}},
  \bibinfo {author} {\bibfnamefont {A.~G.}\ \bibnamefont {Davies}}, \bibinfo
  {author} {\bibfnamefont {E.~H.}\ \bibnamefont {Linfield}},\ and\ \bibinfo
  {author} {\bibfnamefont {R.}~\bibnamefont {Colombelli}},\ }\bibfield  {title}
  {\bibinfo {title} {Ultrafast terahertz detectors based on three-dimensional
  meta-atoms},\ }\href {https://doi.org/10.1364/OPTICA.4.001451} {\bibfield
  {journal} {\bibinfo  {journal} {Optica}\ }\textbf {\bibinfo {volume} {4}},\
  \bibinfo {pages} {1451} (\bibinfo {year} {2017})}\BibitemShut {NoStop}%
\bibitem [{\citenamefont {Jeannin}\ \emph {et~al.}(2019)\citenamefont
  {Jeannin}, \citenamefont {Nesurini}, \citenamefont {Suffit}, \citenamefont
  {Gacemi}, \citenamefont {Vasanelli}, \citenamefont {Li}, \citenamefont
  {Davies}, \citenamefont {Linfield}, \citenamefont {Sirtori},\ and\
  \citenamefont {Todorov}}]{Jeannin2019}%
  \BibitemOpen
  \bibfield  {author} {\bibinfo {author} {\bibfnamefont {M.}~\bibnamefont
  {Jeannin}}, \bibinfo {author} {\bibfnamefont {G.~M.}\ \bibnamefont
  {Nesurini}}, \bibinfo {author} {\bibfnamefont {S.}~\bibnamefont {Suffit}},
  \bibinfo {author} {\bibfnamefont {D.}~\bibnamefont {Gacemi}}, \bibinfo
  {author} {\bibfnamefont {A.}~\bibnamefont {Vasanelli}}, \bibinfo {author}
  {\bibfnamefont {L.}~\bibnamefont {Li}}, \bibinfo {author} {\bibfnamefont
  {A.~G.}\ \bibnamefont {Davies}}, \bibinfo {author} {\bibfnamefont
  {E.}~\bibnamefont {Linfield}}, \bibinfo {author} {\bibfnamefont
  {C.}~\bibnamefont {Sirtori}},\ and\ \bibinfo {author} {\bibfnamefont
  {Y.}~\bibnamefont {Todorov}},\ }\bibfield  {title} {\bibinfo {title}
  {Ultrastrong light{\textendash}matter coupling in deeply subwavelength {THz}
  {LC} resonators},\ }\href {https://doi.org/10.1021/acsphotonics.8b01778}
  {\bibfield  {journal} {\bibinfo  {journal} {{ACS} Photonics}\ }\textbf
  {\bibinfo {volume} {6}},\ \bibinfo {pages} {1207} (\bibinfo {year}
  {2019})}\BibitemShut {NoStop}%
\bibitem [{\citenamefont {Kockum}\ \emph {et~al.}(2019)\citenamefont {Kockum},
  \citenamefont {Miranowicz}, \citenamefont {Liberato}, \citenamefont
  {Savasta},\ and\ \citenamefont {Nori}}]{FriskKockum2019}%
  \BibitemOpen
  \bibfield  {author} {\bibinfo {author} {\bibfnamefont {A.~F.}\ \bibnamefont
  {Kockum}}, \bibinfo {author} {\bibfnamefont {A.}~\bibnamefont {Miranowicz}},
  \bibinfo {author} {\bibfnamefont {S.~D.}\ \bibnamefont {Liberato}}, \bibinfo
  {author} {\bibfnamefont {S.}~\bibnamefont {Savasta}},\ and\ \bibinfo {author}
  {\bibfnamefont {F.}~\bibnamefont {Nori}},\ }\bibfield  {title} {\bibinfo
  {title} {Ultrastrong coupling between light and matter},\ }\href
  {https://doi.org/10.1038/s42254-018-0006-2} {\bibfield  {journal} {\bibinfo
  {journal} {Nature Reviews Physics}\ }\textbf {\bibinfo {volume} {1}},\
  \bibinfo {pages} {19} (\bibinfo {year} {2019})}\BibitemShut {NoStop}%
\bibitem [{\citenamefont {Forn-D\'{\i}az}\ \emph {et~al.}(2019)\citenamefont
  {Forn-D\'{\i}az}, \citenamefont {Lamata}, \citenamefont {Rico}, \citenamefont
  {Kono},\ and\ \citenamefont {Solano}}]{RevModPhys.91.025005}%
  \BibitemOpen
  \bibfield  {author} {\bibinfo {author} {\bibfnamefont {P.}~\bibnamefont
  {Forn-D\'{\i}az}}, \bibinfo {author} {\bibfnamefont {L.}~\bibnamefont
  {Lamata}}, \bibinfo {author} {\bibfnamefont {E.}~\bibnamefont {Rico}},
  \bibinfo {author} {\bibfnamefont {J.}~\bibnamefont {Kono}},\ and\ \bibinfo
  {author} {\bibfnamefont {E.}~\bibnamefont {Solano}},\ }\bibfield  {title}
  {\bibinfo {title} {Ultrastrong coupling regimes of light-matter
  interaction},\ }\href {https://doi.org/10.1103/RevModPhys.91.025005}
  {\bibfield  {journal} {\bibinfo  {journal} {Rev. Mod. Phys.}\ }\textbf
  {\bibinfo {volume} {91}},\ \bibinfo {pages} {025005} (\bibinfo {year}
  {2019})}\BibitemShut {NoStop}%
\bibitem [{\citenamefont {Ciuti}(2021)}]{Ciuti2021}%
  \BibitemOpen
  \bibfield  {author} {\bibinfo {author} {\bibfnamefont {C.}~\bibnamefont
  {Ciuti}},\ }\bibfield  {title} {\bibinfo {title} {Cavity-mediated electron
  hopping in disordered quantum {H}all systems},\ }\href
  {https://doi.org/10.1103/PhysRevB.104.155307} {\bibfield  {journal} {\bibinfo
   {journal} {Phys. Rev. B}\ }\textbf {\bibinfo {volume} {104}},\ \bibinfo
  {pages} {155307} (\bibinfo {year} {2021})}\BibitemShut {NoStop}%
\bibitem [{\citenamefont {Malrieu}\ \emph {et~al.}(1985)\citenamefont
  {Malrieu}, \citenamefont {Durand},\ and\ \citenamefont
  {Daudey}}]{Malrieu_1985}%
  \BibitemOpen
  \bibfield  {author} {\bibinfo {author} {\bibfnamefont {J.~P.}\ \bibnamefont
  {Malrieu}}, \bibinfo {author} {\bibfnamefont {P.}~\bibnamefont {Durand}},\
  and\ \bibinfo {author} {\bibfnamefont {J.~P.}\ \bibnamefont {Daudey}},\
  }\bibfield  {title} {\bibinfo {title} {Intermediate {H}amiltonians as a new
  class of effective {H}amiltonians},\ }\href
  {https://doi.org/10.1088/0305-4470/18/5/014} {\bibfield  {journal} {\bibinfo
  {journal} {Journal of Physics A: Mathematical and General}\ }\textbf
  {\bibinfo {volume} {18}},\ \bibinfo {pages} {809} (\bibinfo {year}
  {1985})}\BibitemShut {NoStop}%
\bibitem [{\citenamefont {Moreira}\ \emph {et~al.}(2002)\citenamefont
  {Moreira}, \citenamefont {Suaud}, \citenamefont {Guih\'ery}, \citenamefont
  {Malrieu}, \citenamefont {Caballol}, \citenamefont {Bofill},\ and\
  \citenamefont {Illas}}]{Effective_PRB}%
  \BibitemOpen
  \bibfield  {author} {\bibinfo {author} {\bibfnamefont {I.~d. P.~R.}\
  \bibnamefont {Moreira}}, \bibinfo {author} {\bibfnamefont {N.}~\bibnamefont
  {Suaud}}, \bibinfo {author} {\bibfnamefont {N.}~\bibnamefont {Guih\'ery}},
  \bibinfo {author} {\bibfnamefont {J.~P.}\ \bibnamefont {Malrieu}}, \bibinfo
  {author} {\bibfnamefont {R.}~\bibnamefont {Caballol}}, \bibinfo {author}
  {\bibfnamefont {J.~M.}\ \bibnamefont {Bofill}},\ and\ \bibinfo {author}
  {\bibfnamefont {F.}~\bibnamefont {Illas}},\ }\bibfield  {title} {\bibinfo
  {title} {Derivation of spin {H}amiltonians from the exact {H}amiltonian:
  Application to systems with two unpaired electrons per magnetic site},\
  }\href {https://doi.org/10.1103/PhysRevB.66.134430} {\bibfield  {journal}
  {\bibinfo  {journal} {Phys. Rev. B}\ }\textbf {\bibinfo {volume} {66}},\
  \bibinfo {pages} {134430} (\bibinfo {year} {2002})}\BibitemShut {NoStop}%
\bibitem [{\citenamefont {Datta}(2005)}]{Datta2005}%
  \BibitemOpen
  \bibfield  {author} {\bibinfo {author} {\bibfnamefont {S.}~\bibnamefont
  {Datta}},\ }\href {https://doi.org/10.1017/cbo9781139164313} {\emph {\bibinfo
  {title} {Quantum Transport: Atom to Transistor}}}\ (\bibinfo  {publisher}
  {Cambridge University Press},\ \bibinfo {year} {2005})\BibitemShut {NoStop}%
\bibitem [{\citenamefont {Girvin}\ and\ \citenamefont
  {Yang}(2019)}]{girvin_yang_2019}%
  \BibitemOpen
  \bibfield  {author} {\bibinfo {author} {\bibfnamefont {S.~M.}\ \bibnamefont
  {Girvin}}\ and\ \bibinfo {author} {\bibfnamefont {K.}~\bibnamefont {Yang}},\
  }\href {https://doi.org/10.1017/9781316480649} {\emph {\bibinfo {title}
  {Modern Condensed Matter Physics}}}\ (\bibinfo  {publisher} {Cambridge
  University Press},\ \bibinfo {year} {2019})\BibitemShut {NoStop}%
\bibitem [{\citenamefont {van Wees}\ \emph {et~al.}(1988)\citenamefont {van
  Wees}, \citenamefont {van Houten}, \citenamefont {Beenakker}, \citenamefont
  {Williamson}, \citenamefont {Kouwenhoven}, \citenamefont {van~der Marel},\
  and\ \citenamefont {Foxon}}]{QPC_1988}%
  \BibitemOpen
  \bibfield  {author} {\bibinfo {author} {\bibfnamefont {B.~J.}\ \bibnamefont
  {van Wees}}, \bibinfo {author} {\bibfnamefont {H.}~\bibnamefont {van
  Houten}}, \bibinfo {author} {\bibfnamefont {C.~W.~J.}\ \bibnamefont
  {Beenakker}}, \bibinfo {author} {\bibfnamefont {J.~G.}\ \bibnamefont
  {Williamson}}, \bibinfo {author} {\bibfnamefont {L.~P.}\ \bibnamefont
  {Kouwenhoven}}, \bibinfo {author} {\bibfnamefont {D.}~\bibnamefont {van~der
  Marel}},\ and\ \bibinfo {author} {\bibfnamefont {C.~T.}\ \bibnamefont
  {Foxon}},\ }\bibfield  {title} {\bibinfo {title} {Quantized conductance of
  point contacts in a two-dimensional electron gas},\ }\href
  {https://doi.org/10.1103/PhysRevLett.60.848} {\bibfield  {journal} {\bibinfo
  {journal} {Phys. Rev. Lett.}\ }\textbf {\bibinfo {volume} {60}},\ \bibinfo
  {pages} {848} (\bibinfo {year} {1988})}\BibitemShut {NoStop}%
\bibitem [{\citenamefont {Wharam}\ \emph {et~al.}(1988)\citenamefont {Wharam},
  \citenamefont {Thornton}, \citenamefont {Newbury}, \citenamefont {Pepper},
  \citenamefont {Ahmed}, \citenamefont {Frost}, \citenamefont {Hasko},
  \citenamefont {Peacock}, \citenamefont {Ritchie},\ and\ \citenamefont
  {Jones}}]{wharam1988one}%
  \BibitemOpen
  \bibfield  {author} {\bibinfo {author} {\bibfnamefont {D.~A.}\ \bibnamefont
  {Wharam}}, \bibinfo {author} {\bibfnamefont {T.~J.}\ \bibnamefont
  {Thornton}}, \bibinfo {author} {\bibfnamefont {R.}~\bibnamefont {Newbury}},
  \bibinfo {author} {\bibfnamefont {M.}~\bibnamefont {Pepper}}, \bibinfo
  {author} {\bibfnamefont {H.}~\bibnamefont {Ahmed}}, \bibinfo {author}
  {\bibfnamefont {J.}~\bibnamefont {Frost}}, \bibinfo {author} {\bibfnamefont
  {D.}~\bibnamefont {Hasko}}, \bibinfo {author} {\bibfnamefont
  {D.}~\bibnamefont {Peacock}}, \bibinfo {author} {\bibfnamefont
  {D.}~\bibnamefont {Ritchie}},\ and\ \bibinfo {author} {\bibfnamefont
  {G.}~\bibnamefont {Jones}},\ }\bibfield  {title} {\bibinfo {title}
  {One-dimensional transport and the quantisation of the ballistic
  resistance},\ }\href@noop {} {\bibfield  {journal} {\bibinfo  {journal}
  {Journal of Physics C: solid state physics}\ }\textbf {\bibinfo {volume}
  {21}},\ \bibinfo {pages} {L209} (\bibinfo {year} {1988})}\BibitemShut
  {NoStop}%
\bibitem [{\citenamefont {Kane}\ \emph {et~al.}(1998)\citenamefont {Kane},
  \citenamefont {Facer}, \citenamefont {Dzurak}, \citenamefont {Lumpkin},
  \citenamefont {Clark}, \citenamefont {Pfeiffer},\ and\ \citenamefont
  {West}}]{kane1998quantized}%
  \BibitemOpen
  \bibfield  {author} {\bibinfo {author} {\bibfnamefont {B.~E.}\ \bibnamefont
  {Kane}}, \bibinfo {author} {\bibfnamefont {G.~R.}\ \bibnamefont {Facer}},
  \bibinfo {author} {\bibfnamefont {A.~S.}\ \bibnamefont {Dzurak}}, \bibinfo
  {author} {\bibfnamefont {N.~E.}\ \bibnamefont {Lumpkin}}, \bibinfo {author}
  {\bibfnamefont {R.~G.}\ \bibnamefont {Clark}}, \bibinfo {author}
  {\bibfnamefont {L.~N.}\ \bibnamefont {Pfeiffer}},\ and\ \bibinfo {author}
  {\bibfnamefont {K.~W.}\ \bibnamefont {West}},\ }\bibfield  {title} {\bibinfo
  {title} {Quantized conductance in quantum wires with gate-controlled width
  and electron density},\ }\href {https://doi.org/10.1063/1.121642} {\bibfield
  {journal} {\bibinfo  {journal} {Applied Physics Letters}\ }\textbf {\bibinfo
  {volume} {72}},\ \bibinfo {pages} {3506} (\bibinfo {year}
  {1998})}\BibitemShut {NoStop}%
\bibitem [{\citenamefont {Irber}\ \emph {et~al.}(2017)\citenamefont {Irber},
  \citenamefont {Seidl}, \citenamefont {Carrad}, \citenamefont {Becker},
  \citenamefont {Jeon}, \citenamefont {Loitsch}, \citenamefont {Winnerl},
  \citenamefont {Matich}, \citenamefont {D\"{o}blinger}, \citenamefont {Tang},
  \citenamefont {Mork\"{o}tter}, \citenamefont {Abstreiter}, \citenamefont
  {Finley}, \citenamefont {Grayson}, \citenamefont {Lauhon},\ and\
  \citenamefont {Koblm\"{u}ller}}]{irber2017quantum}%
  \BibitemOpen
  \bibfield  {author} {\bibinfo {author} {\bibfnamefont {D.~M.}\ \bibnamefont
  {Irber}}, \bibinfo {author} {\bibfnamefont {J.}~\bibnamefont {Seidl}},
  \bibinfo {author} {\bibfnamefont {D.~J.}\ \bibnamefont {Carrad}}, \bibinfo
  {author} {\bibfnamefont {J.}~\bibnamefont {Becker}}, \bibinfo {author}
  {\bibfnamefont {N.}~\bibnamefont {Jeon}}, \bibinfo {author} {\bibfnamefont
  {B.}~\bibnamefont {Loitsch}}, \bibinfo {author} {\bibfnamefont
  {J.}~\bibnamefont {Winnerl}}, \bibinfo {author} {\bibfnamefont
  {S.}~\bibnamefont {Matich}}, \bibinfo {author} {\bibfnamefont
  {M.}~\bibnamefont {D\"{o}blinger}}, \bibinfo {author} {\bibfnamefont
  {Y.}~\bibnamefont {Tang}}, \bibinfo {author} {\bibfnamefont {S.}~\bibnamefont
  {Mork\"{o}tter}}, \bibinfo {author} {\bibfnamefont {G.}~\bibnamefont
  {Abstreiter}}, \bibinfo {author} {\bibfnamefont {J.~J.}\ \bibnamefont
  {Finley}}, \bibinfo {author} {\bibfnamefont {M.}~\bibnamefont {Grayson}},
  \bibinfo {author} {\bibfnamefont {L.~J.}\ \bibnamefont {Lauhon}},\ and\
  \bibinfo {author} {\bibfnamefont {G.}~\bibnamefont {Koblm\"{u}ller}},\
  }\bibfield  {title} {\bibinfo {title} {Quantum transport and sub-band
  structure of modulation-doped {GaAs}/{AlAs} core{\textendash}superlattice
  nanowires},\ }\href {https://doi.org/10.1021/acs.nanolett.7b01732} {\bibfield
   {journal} {\bibinfo  {journal} {Nano Letters}\ }\textbf {\bibinfo {volume}
  {17}},\ \bibinfo {pages} {4886} (\bibinfo {year} {2017})}\BibitemShut
  {NoStop}%
\bibitem [{\citenamefont {Keller}\ \emph {et~al.}(2017)\citenamefont {Keller},
  \citenamefont {Scalari}, \citenamefont {Cibella}, \citenamefont {Maissen},
  \citenamefont {Appugliese}, \citenamefont {Giovine}, \citenamefont {Leoni},
  \citenamefont {Beck},\ and\ \citenamefont {Faist}}]{keller2017few}%
  \BibitemOpen
  \bibfield  {author} {\bibinfo {author} {\bibfnamefont {J.}~\bibnamefont
  {Keller}}, \bibinfo {author} {\bibfnamefont {G.}~\bibnamefont {Scalari}},
  \bibinfo {author} {\bibfnamefont {S.}~\bibnamefont {Cibella}}, \bibinfo
  {author} {\bibfnamefont {C.}~\bibnamefont {Maissen}}, \bibinfo {author}
  {\bibfnamefont {F.}~\bibnamefont {Appugliese}}, \bibinfo {author}
  {\bibfnamefont {E.}~\bibnamefont {Giovine}}, \bibinfo {author} {\bibfnamefont
  {R.}~\bibnamefont {Leoni}}, \bibinfo {author} {\bibfnamefont
  {M.}~\bibnamefont {Beck}},\ and\ \bibinfo {author} {\bibfnamefont
  {J.}~\bibnamefont {Faist}},\ }\bibfield  {title} {\bibinfo {title}
  {Few-electron ultrastrong light-matter coupling at 300 {GH}z with nanogap
  hybrid {LC} microcavities},\ }\href
  {https://doi.org/https://doi.org/10.1021/acs.nanolett.7b03228} {\bibfield
  {journal} {\bibinfo  {journal} {Nano letters}\ }\textbf {\bibinfo {volume}
  {17}},\ \bibinfo {pages} {7410} (\bibinfo {year} {2017})}\BibitemShut
  {NoStop}%
\bibitem [{\citenamefont {B\"uttiker}(1988)}]{Buttiker1988}%
  \BibitemOpen
  \bibfield  {author} {\bibinfo {author} {\bibfnamefont {M.}~\bibnamefont
  {B\"uttiker}},\ }\bibfield  {title} {\bibinfo {title} {Absence of
  backscattering in the quantum {H}all effect in multiprobe conductors},\
  }\href {https://doi.org/10.1103/PhysRevB.38.9375} {\bibfield  {journal}
  {\bibinfo  {journal} {Phys. Rev. B}\ }\textbf {\bibinfo {volume} {38}},\
  \bibinfo {pages} {9375} (\bibinfo {year} {1988})}\BibitemShut {NoStop}%
\bibitem [{\citenamefont {Baranger}\ and\ \citenamefont
  {Stone}(1989)}]{Baranger1989}%
  \BibitemOpen
  \bibfield  {author} {\bibinfo {author} {\bibfnamefont {H.~U.}\ \bibnamefont
  {Baranger}}\ and\ \bibinfo {author} {\bibfnamefont {A.~D.}\ \bibnamefont
  {Stone}},\ }\bibfield  {title} {\bibinfo {title} {Electrical linear-response
  theory in an arbitrary magnetic field: A new {F}ermi-surface formation},\
  }\href {https://doi.org/10.1103/PhysRevB.40.8169} {\bibfield  {journal}
  {\bibinfo  {journal} {Phys. Rev. B}\ }\textbf {\bibinfo {volume} {40}},\
  \bibinfo {pages} {8169} (\bibinfo {year} {1989})}\BibitemShut {NoStop}%
\bibitem [{Note1()}]{Note1}%
  \BibitemOpen
  \bibinfo {note} {With the cavity-mediated effective Hamiltonian, the
  calculation is heavier because the matrix $\protect \mathaccentV
  {tilde}07E{\Gamma }_{\lambda ,\lambda '}$ is not sparse and describes an
  all-to-all coupling between the bare eigenstates. Correspondingly, also the
  calculation of the transmission coefficients is more demanding.}\BibitemShut
  {Stop}%
\bibitem [{\citenamefont {Dean}\ \emph {et~al.}(2013)\citenamefont {Dean},
  \citenamefont {Wang}, \citenamefont {Maher}, \citenamefont {Forsythe},
  \citenamefont {Ghahari}, \citenamefont {Gao}, \citenamefont {Katoch},
  \citenamefont {Ishigami}, \citenamefont {Moon}, \citenamefont {Koshino} \emph
  {et~al.}}]{dean2013hofstadter}%
  \BibitemOpen
  \bibfield  {author} {\bibinfo {author} {\bibfnamefont {C.~R.}\ \bibnamefont
  {Dean}}, \bibinfo {author} {\bibfnamefont {L.}~\bibnamefont {Wang}}, \bibinfo
  {author} {\bibfnamefont {P.}~\bibnamefont {Maher}}, \bibinfo {author}
  {\bibfnamefont {C.}~\bibnamefont {Forsythe}}, \bibinfo {author}
  {\bibfnamefont {F.}~\bibnamefont {Ghahari}}, \bibinfo {author} {\bibfnamefont
  {Y.}~\bibnamefont {Gao}}, \bibinfo {author} {\bibfnamefont {J.}~\bibnamefont
  {Katoch}}, \bibinfo {author} {\bibfnamefont {M.}~\bibnamefont {Ishigami}},
  \bibinfo {author} {\bibfnamefont {P.}~\bibnamefont {Moon}}, \bibinfo {author}
  {\bibfnamefont {M.}~\bibnamefont {Koshino}}, \emph {et~al.},\ }\bibfield
  {title} {\bibinfo {title} {Hofstadter’s butterfly and the fractal quantum
  {H}all effect in moir{\'e} superlattices},\ }\href
  {https://www.nature.com/articles/nature12186} {\bibfield  {journal} {\bibinfo
   {journal} {Nature}\ }\textbf {\bibinfo {volume} {497}},\ \bibinfo {pages}
  {598} (\bibinfo {year} {2013})}\BibitemShut {NoStop}%
\bibitem [{\citenamefont {Rokaj}\ \emph {et~al.}(2022)\citenamefont {Rokaj},
  \citenamefont {Penz}, \citenamefont {Sentef}, \citenamefont {Ruggenthaler},\
  and\ \citenamefont {Rubio}}]{Rubio2022}%
  \BibitemOpen
  \bibfield  {author} {\bibinfo {author} {\bibfnamefont {V.}~\bibnamefont
  {Rokaj}}, \bibinfo {author} {\bibfnamefont {M.}~\bibnamefont {Penz}},
  \bibinfo {author} {\bibfnamefont {M.~A.}\ \bibnamefont {Sentef}}, \bibinfo
  {author} {\bibfnamefont {M.}~\bibnamefont {Ruggenthaler}},\ and\ \bibinfo
  {author} {\bibfnamefont {A.}~\bibnamefont {Rubio}},\ }\bibfield  {title}
  {\bibinfo {title} {Polaritonic {H}ofstadter butterfly and cavity control of
  the quantized {H}all conductance},\ }\href
  {https://doi.org/10.1103/PhysRevB.105.205424} {\bibfield  {journal} {\bibinfo
   {journal} {Phys. Rev. B}\ }\textbf {\bibinfo {volume} {105}},\ \bibinfo
  {pages} {205424} (\bibinfo {year} {2022})}\BibitemShut {NoStop}%
\bibitem [{\citenamefont {Vigneron}\ \emph {et~al.}(2019)\citenamefont
  {Vigneron}, \citenamefont {Pirotta}, \citenamefont {Carusotto}, \citenamefont
  {Tran}, \citenamefont {Biasiol}, \citenamefont {Manceau}, \citenamefont
  {Bousseksou},\ and\ \citenamefont {Colombelli}}]{vigneron2019quantum}%
  \BibitemOpen
  \bibfield  {author} {\bibinfo {author} {\bibfnamefont {P.-B.}\ \bibnamefont
  {Vigneron}}, \bibinfo {author} {\bibfnamefont {S.}~\bibnamefont {Pirotta}},
  \bibinfo {author} {\bibfnamefont {I.}~\bibnamefont {Carusotto}}, \bibinfo
  {author} {\bibfnamefont {N.-L.}\ \bibnamefont {Tran}}, \bibinfo {author}
  {\bibfnamefont {G.}~\bibnamefont {Biasiol}}, \bibinfo {author} {\bibfnamefont
  {J.-M.}\ \bibnamefont {Manceau}}, \bibinfo {author} {\bibfnamefont
  {A.}~\bibnamefont {Bousseksou}},\ and\ \bibinfo {author} {\bibfnamefont
  {R.}~\bibnamefont {Colombelli}},\ }\bibfield  {title} {\bibinfo {title}
  {Quantum well infrared photo-detectors operating in the strong light-matter
  coupling regime},\ }\href
  {https://aip.scitation.org/doi/full/10.1063/1.5084112} {\bibfield  {journal}
  {\bibinfo  {journal} {Applied Physics Letters}\ }\textbf {\bibinfo {volume}
  {114}},\ \bibinfo {pages} {131104} (\bibinfo {year} {2019})}\BibitemShut
  {NoStop}%
\bibitem [{\citenamefont {Limbacher}\ \emph {et~al.}(2020)\citenamefont
  {Limbacher}, \citenamefont {Kainz}, \citenamefont {Schoenhuber},
  \citenamefont {Wenclawiak}, \citenamefont {Derntl}, \citenamefont {Andrews},
  \citenamefont {Detz}, \citenamefont {Strasser}, \citenamefont {Schwaighofer},
  \citenamefont {Lendl}, \citenamefont {Darmo},\ and\ \citenamefont
  {Unterrainer}}]{Limbacher2020}%
  \BibitemOpen
  \bibfield  {author} {\bibinfo {author} {\bibfnamefont {B.}~\bibnamefont
  {Limbacher}}, \bibinfo {author} {\bibfnamefont {M.~A.}\ \bibnamefont
  {Kainz}}, \bibinfo {author} {\bibfnamefont {S.}~\bibnamefont {Schoenhuber}},
  \bibinfo {author} {\bibfnamefont {M.}~\bibnamefont {Wenclawiak}}, \bibinfo
  {author} {\bibfnamefont {C.}~\bibnamefont {Derntl}}, \bibinfo {author}
  {\bibfnamefont {A.~M.}\ \bibnamefont {Andrews}}, \bibinfo {author}
  {\bibfnamefont {H.}~\bibnamefont {Detz}}, \bibinfo {author} {\bibfnamefont
  {G.}~\bibnamefont {Strasser}}, \bibinfo {author} {\bibfnamefont
  {A.}~\bibnamefont {Schwaighofer}}, \bibinfo {author} {\bibfnamefont
  {B.}~\bibnamefont {Lendl}}, \bibinfo {author} {\bibfnamefont
  {J.}~\bibnamefont {Darmo}},\ and\ \bibinfo {author} {\bibfnamefont
  {K.}~\bibnamefont {Unterrainer}},\ }\bibfield  {title} {\bibinfo {title}
  {Resonant tunneling diodes strongly coupled to the cavity field},\ }\href
  {https://doi.org/10.1063/5.0007118} {\bibfield  {journal} {\bibinfo
  {journal} {Applied Physics Letters}\ }\textbf {\bibinfo {volume} {116}},\
  \bibinfo {pages} {221101} (\bibinfo {year} {2020})}\BibitemShut {NoStop}%
\bibitem [{\citenamefont {Hatefipour}\ \emph {et~al.}(2021)\citenamefont
  {Hatefipour}, \citenamefont {Cuozzo}, \citenamefont {Kanter}, \citenamefont
  {Strickland}, \citenamefont {Lu}, \citenamefont {Rossi},\ and\ \citenamefont
  {Shabani}}]{hatefipour2021induced}%
  \BibitemOpen
  \bibfield  {author} {\bibinfo {author} {\bibfnamefont {M.}~\bibnamefont
  {Hatefipour}}, \bibinfo {author} {\bibfnamefont {J.~J.}\ \bibnamefont
  {Cuozzo}}, \bibinfo {author} {\bibfnamefont {J.}~\bibnamefont {Kanter}},
  \bibinfo {author} {\bibfnamefont {W.}~\bibnamefont {Strickland}}, \bibinfo
  {author} {\bibfnamefont {T.-M.}\ \bibnamefont {Lu}}, \bibinfo {author}
  {\bibfnamefont {E.}~\bibnamefont {Rossi}},\ and\ \bibinfo {author}
  {\bibfnamefont {J.}~\bibnamefont {Shabani}},\ }\bibfield  {title} {\bibinfo
  {title} {Induced superconducting pairing in integer quantum {H}all edge
  states},\ }\href {https://doi.org/10.48550/arXiv.2108.08899} {\bibfield
  {journal} {\bibinfo  {journal} {arXiv preprint arXiv:2108.08899}\ } (\bibinfo
  {year} {2021})}\BibitemShut {NoStop}%
\bibitem [{\citenamefont {Groth}\ \emph {et~al.}(2014)\citenamefont {Groth},
  \citenamefont {Wimmer}, \citenamefont {Akhmerov},\ and\ \citenamefont
  {Waintal}}]{Kwant}%
  \BibitemOpen
  \bibfield  {author} {\bibinfo {author} {\bibfnamefont {C.~W.}\ \bibnamefont
  {Groth}}, \bibinfo {author} {\bibfnamefont {M.}~\bibnamefont {Wimmer}},
  \bibinfo {author} {\bibfnamefont {A.~R.}\ \bibnamefont {Akhmerov}},\ and\
  \bibinfo {author} {\bibfnamefont {X.}~\bibnamefont {Waintal}},\ }\bibfield
  {title} {\bibinfo {title} {Kwant: a software package for quantum transport},\
  }\href {https://doi.org/10.1088/1367-2630/16/6/063065} {\bibfield  {journal}
  {\bibinfo  {journal} {New Journal of Physics}\ }\textbf {\bibinfo {volume}
  {16}},\ \bibinfo {pages} {063065} (\bibinfo {year} {2014})}\BibitemShut
  {NoStop}%
\end{thebibliography}%

\end{document}